\documentclass{scrartcl}

\usepackage{graphicx}

\DeclareRobustCommand{\pluralism}{\raisebox{-0.1em}{\includegraphics[height=.9em]{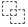}}\ }

\DeclareRobustCommand{\nameraka}{\raisebox{-0.1em}{\includegraphics[height=.9em]{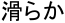}}}

\usepackage[semibold]{inter}

\usepackage{booktabs}
\usepackage{array}

\usepackage[style=authoryear, backend=biber, uniquename=false]{biblatex}
\usepackage[colorlinks=false, pdfborder={0 0 0}]{hyperref}

\usepackage{xcolor}
\definecolor{muted}{HTML}{777777}

\let\cite\parencite

\usepackage{amssymb}

\newcommand\footertext[1]{%
  \begingroup
  \renewcommand\thefootnote{}\footnote{#1}%
  \addtocounter{footnote}{-1}%
  \endgroup
}

\usepackage[T1]{fontenc}
\usepackage{ebgaramond}

\addtokomafont{disposition}{\rmfamily\normalfont}
\setkomafont{paragraph}{\itshape}

\title{Community by Design}

\author{
    E. Glen Weyl, 
    Luke Thorburn, 
    Emillie de Keulenaar,\\  
    Jacob Mchangama, 
    Divya Siddarth, and
    Audrey Tang
}

\date{}

\begin{document}

\maketitle

\begin{abstract}
    Social media empower distributed content creation by algorithmically harnessing ``the social fabric'' (explicit and implicit signals of association) to serve this content.  While this overcomes the bottlenecks and biases of traditional gatekeepers, many believe it has unsustainably eroded the very social fabric it depends on by maximizing engagement for advertising revenue.  This paper participates in open and ongoing considerations to translate social and political values and conventions — specifically social cohesion — into platform design. We propose an alternative platform model that includes a formalization of the social fabric as an explicit output as well as input.  In this formalization, citizens are members of communities defined by explicit affiliation or clusters of shared attitudes.  Both have internal divisions, as citizens are members of intersecting communities, which are themselves internally diverse.  Each is understood to value content that bridge (viz. achieve consensus across) and balance (viz. represent fairly) this internal diversity, consistent with the principles of the Hutchins Commission (1947).  Content is labeled with social provenance, indicating for which community or citizen it is bridging or balancing.  Subscription payments allow citizens and communities to increase the algorithmic weight on the content they value in the content serving algorithm.  Advertisers may, with consent of citizen or community counterparties, target them in exchange for payment or increase in that party's algorithmic weight.  Underserved and emerging communities and citizens are optimally subsidized/supported to develop into paying participants.  Content creators and communities that curate content are rewarded for their contributions with algorithmic weight and/or revenue.  We discuss applications to productivity (e.g. LinkedIn), political (e.g. X), and cultural (e.g. TikTok) platforms.
\end{abstract}

\footertext{%
    \textbf{Authors}: E. Glen Weyl, Microsoft Research Special Projects (glenweyl@microsoft.com); Luke Thorburn, King's College London (mail@lukethorburn.com); Emillie de Keulenaar, University of Groningen  (e.v.de.keulenaar@rug.nl); Jacob Mchangama, Vanderbilt University (jacob@futurefreespeech.org); Divya Siddarth, Collective Intelligence Project (divya@cip.org); Audrey Tang, Project Liberty Institute (audreyt@audreyt.org).%
}

\footertext{%
    \textbf{Acknowledgements}: We are grateful to Daron Acemoglu, Robin Burke, Nick Couldry, Madeleine Daepp, Mona Hamdy, Jaron Lanier, Puja Ohlhaver, Manon Revel, Jonathan Stray, Jordan Usdan, Chris White, and conference participants at the Project Liberty Summit for their feedback.%
}

\section{Introduction}
\label{sec:intro}

    Social media have transformed communications by replacing editorial curation with algorithms that harness data from the fabric of social interactions to serve content.  Many \cite{lanier2018b,orlowski2020,haidt2024} have argued, however, that the most prominent existing examples of such media are ``antisocial'' \cite{vaidhyanathan2018} because, in the process of harnessing social fabric data to maximize engagement with revenue-generating advertisements, they undermine the social fabric on which they depend.  Most existing attempts to address this dynamic involve top-down content moderation that often fails to address the problem systemically, be that in marginalized communities that receive less top-down attention \cite{suzor2019} or by challenging principles of free speech while centralizing editorial control in the hands of unaccountable technical elites \cite{gillespie2018, west2018}.
    
    To overcome this dilemma, we propose a communitarian platform design model based on formalizing the social fabric as a hypergraph \cite{simmel1908}, with each edge (viz.\ community) defined by organization around points of common knowledge \cite{lewis1969,aumann1976} as measured by bridging and balancing content in the spirit of the \textcite{Hutchins} and as articulated by \textcite{ovadya2023a}. Communities and citizens that are members of many of these would pay subscription fees or accept targeted advertising to increase the prevalence of such content for them and be rewarded for content achieving these goals for others.

    While ``social media'' has a variety of definitions, we focus on the diversity of content creators (often called ``user-generated content'') and the replacement \cite{jenkinsCulturalLogicMedia2004} of professional editorial curation with the algorithmic processing of implicit and explicit signals of social affiliation to serve content (often called ``algorithmic curation''). Social media thus conceived brought two critical and closely related benefits.  First, it complemented and sometimes dislodged the ``bottleneck'' role of ``gatekeepers'', typically from a narrow set of social backgrounds in limited geographies, that often reduced the diversity and creativity of content, offering a broad set of pathways to visibility.  Second, it allowed a diversification of audiences beyond those that could afford the fixed costs of editorial curation, publication, hosting and distribution.

    This allowed many previously underserved communities and creators to find a voice and space, and these benefits stretch far beyond the cultural and political spheres in which they are most often discussed: productivity platforms such as LinkedIn and GitHub have similarly helped organizations and industries ``break silos'' created by divisional and corporate structures to increase innovation and speed development. Social media have also allowed citizens in authoritarian states to destabilize official censorship and propaganda by amplifying dissent and coordinating protest movements that undermined and sometimes dislodged entrenched rulers \cite{mcgarty}.

    Yet, as instantiated in practice by leading ``Web 2.0'' platforms \cite{o2009web} such as Facebook, TikTok, and X (formerly Twitter), social media are now widely understood as having created a ``social dilemma'' \cite{orlowski2020}, for while they rely on and harness the rich information embedded in the social fabric, they fail to regenerate and often in fact degrade that fabric.  For one, the dramatic expansion of any and all political voices on social media, while positive in most respects, has not been met with adequate means to facilitate dialogue and understanding across radical differences. Combined with a drive for engagement, this has resulted in profound social division and polarization \cite{klein2020}.  Second, social media are blamed as being primary loci for the spread of mis- and disinformation \cite{vosoughi2018}, partly as a result of poor consensus-building mechanisms.  They are thus seen as central points of vulnerability for adversaries who seek to undermine the common understanding that is the foundation of collective action, especially in pluralistic societies where media is not centralized \cite{king2013,king2017,farrell2018}.  Ironically, social media itself may be centralized, in the sense that it has recreated choke points for control over speech \cite{klonick2017new}, either granting unchecked powers to corporate interest or opportunities for governments to suppress dissent \cite{ffs2023,ffs2024}. Finally, they are seen as driving a growing epidemic of anxiety and social isolation, especially among young people \cite{haidt2024, couldry2024}.

    Most platform mechanisms look for incentives to maximize engagement by creating strong (and often negative) emotions driven by a personalized advertising business model \cite{wu2017,sunstein2018,zuboff2019}.  While it has clearly proven a natural and profitable business model, this model predates social media and flourished just as well in the ``high modernist'' media era against which present challenges are often contrasted, including the highly polarized American media environment before the Second World War.  The more fundamental problem and distinguishing feature of social media may be one anticipated by \textcite{stephenson1995} in his classic novel \textit{The Diamond Age}, written at the dawn of the internet era.  There, the poor receive personalized and entertaining news feeds, fragmenting their understanding of reality while absorbing their attention, thus undermining their capacity for common action, while the powerful all receive a common authoritative paper \textit{The Times}, which creates a shared culture allowing them to rule.

    This highlights that current platforms have failed to deliver the other half of the equation: the coherence and common understanding that empower collective decision-making, effective social action, cultures of belonging, and so on.  While ``antisocial'' media platforms harness social fabric data, they simultaneously obscure it: people receive content because of the social communities they are inferred to belong to without a clear understanding of the community of others that are seeing and assenting to that content.  They thus can easily come to believe ``viral'' content they see is broadly consensual, when it only reflects the views of a narrow community, fueling the ``false consensus effect'' that is known to underlie polarization and the spread of mis- and disinformation \cite{ecker2022}.  They are easily vulnerable to outside attacks that aim to use social cleavages to undermine coordination \cite{usdoj2024}.  And they may experience various kinds of social isolation, as they lack a sense for common experience and culture with identifiable others \cite{lim2020}.

    But just because these problems are modern does not mean we have to reinvent the wheel. Luckily, we already stand behind a rich and long history of public design --- that is, norms, conventions, frameworks or methods for sustaining social concord in public spaces, be that a town square, a parliament, a workspace or media environments. One such framework came from a media reform movement in the wake of the Second World War and focused precisely on providing such coherence, laying some of the foundation for the high modernist era of prestigious gatekeepers.  In 1942, immediately following US entry into the War, media leaders feared press freedom would become another casualty given the way in which the highly polarized media landscape in the run up to Pearl Harbor had fueled isolationism and undermined national morale.  To preempt such a possibility, \textit{TIME Magazine}'s Henry Luce asked University of Chicago President Robert Hutchins to convene a commission on press freedom and responsibility to lay out standards of professionalism that were to define the US industry and its regulation in the following decades.  Their 1947 report laid out five principles we discuss below, but can be briefly summarized as: 1) drawing a distinction between ``news'' and ``opinion''; 2) defining news as things that can be agreed across (viz. bridge) a diverse society; and 3) ensuring that opinion that divides that society is represented with balance.  For much of the postwar period, implementing the Hutchins principles was the central responsibility of editorial staff.  Yet, combined with other international media principles, they have also influenced editorial guidelines in the information age — for example, the combination of a ``neutral point of view'' doctrine with robust consensus-building mechanisms on Wikipedia \cite{matei2011}.

\begin{figure}[t]
        \centering
        \includegraphics[width=\linewidth]{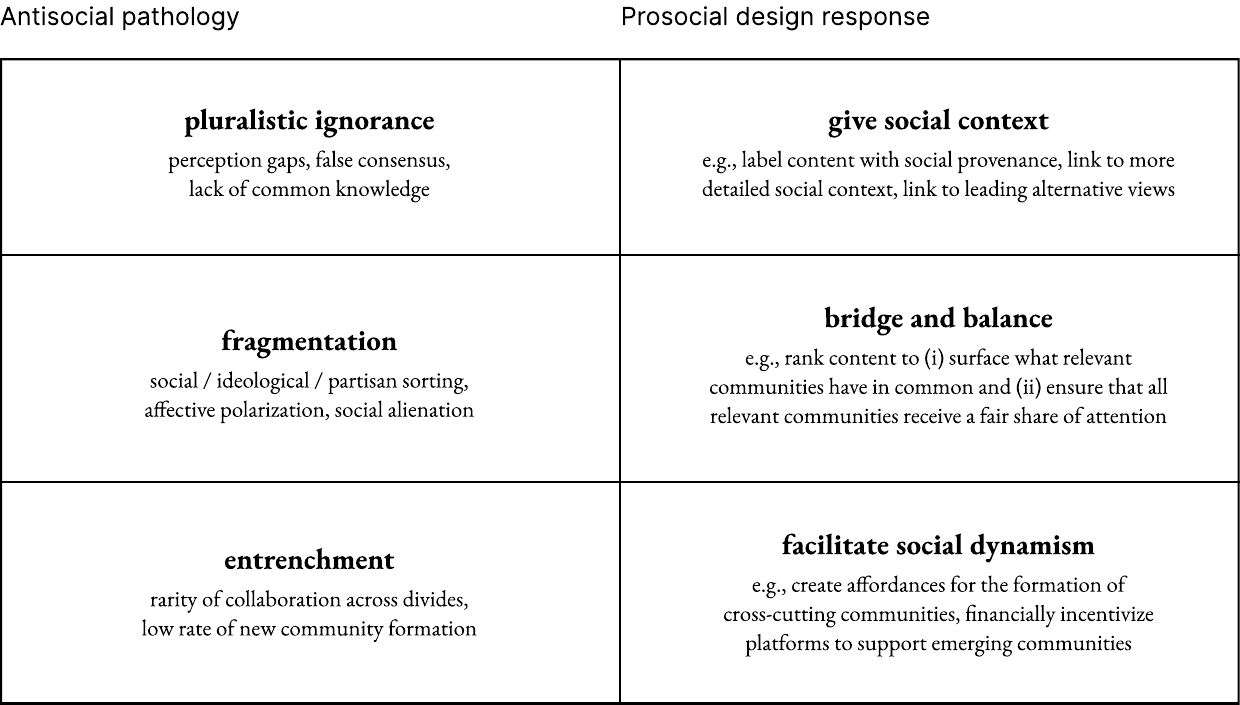}
        \caption{Summary of three broad, antisocial pathologies of the public sphere, with our corresponding recommendations for the design of social media.}
        \label{fig:interventions}
    \end{figure}

\begin{figure}[t]
        \centering
        \includegraphics[width=\linewidth]{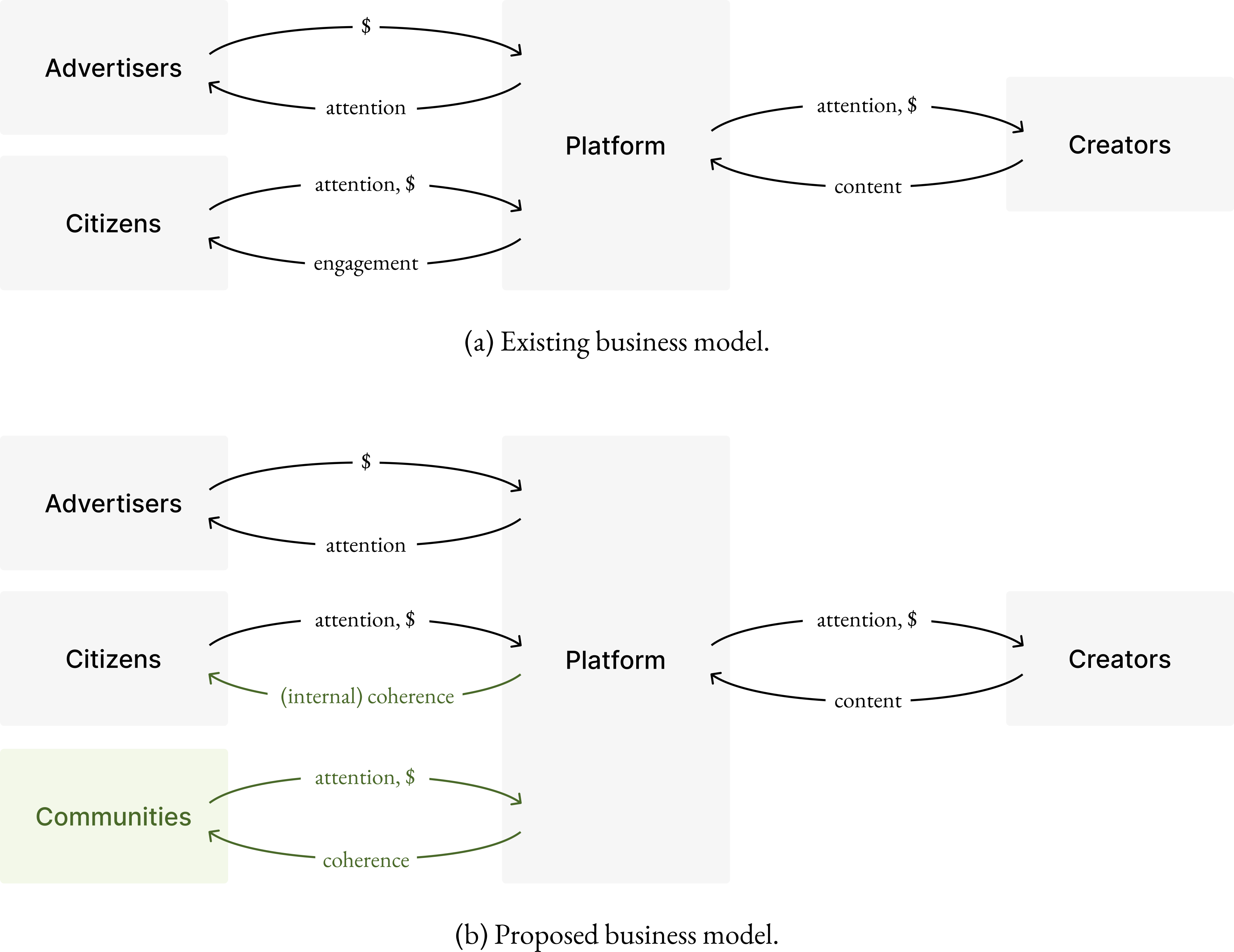}
        \caption{Illustration of (a) the core business model for (most) existing social media platforms, and (b) our proposed business model.}
        \label{fig:business-models}
    \end{figure}

    This paper proposes design features for social media based in ``coherent pluralism'' or ``plurality'', which we encode in the character \pluralism. Our intention is not to be highly prescriptive, but to illustrate an alternative paradigm for social media with sufficient detail that those involved in platform design can apply the core ideas in their context. Concretely, we recommend three kinds design intervention (Figure \ref{fig:interventions}). First, to mitigate pluralistic ignorance and to facilitate the production of common understanding or ``meta-consensus'' \cite{dryzek2006reconciling} that undergirds the social fabric, we recommend providing \textit{social context}, such as by annotating posts with the communities among which they are widely accepted or divisive. Second, to address social fragmentation, we propose ranking content to (i) surface what relevant communities have in common, and (ii) ensure that all relevant communities receive a fair share of attention. We envision this by harnessing and scaling an algorithmic implementation of the principles of the latter proposed by \textcite{faridani2010} and prominently implemented by \textcite{small2021} and \textcite{wojcik2022}. Third, to address entrenchment of social dynamics and the rarity of collaboration across divides, we propose affordances for the formation of cross-cutting communities, and propose a business model (Figure \ref{fig:business-models}) under which platforms are incentivized to support emerging communities.
    
    Our discussion above focused on the setting of news that is perhaps the most widely discussed and is probably most relevant to platforms such as X. However, the goal of \pluralism is at least as relevant to productivity platforms like LinkedIn and cultural platforms like TikTok.  In productivity contexts, \pluralism is critical to allow businesses, entrepreneurial clusters and international alliances to cooperate across diverse working cultures and navigate disruption across organizational boundaries.  Furthermore, many of these platforms have preexisting commercial relationships with such organizations, facilitating the natural business model of selling organizations prominence of content that coheres their associated community.  For the other extreme of primarily cultural platforms like TikTok, \pluralism can facilitate the formation of collective memory that underpin popular culture in a way that major cultural events, be they concerts, exhibits or even sports, have tended to do in the past. They also offer a natural context in which advertising targeted to these large and self-aware audiences can go beyond the value of personalized advertising to create shared meaning around a consumption experience, as many classic advertising campaigns (e.g. ``share a Coke'' or Apple's 1984) have tried to.

    To construct our proposed design, we draw on technical and theoretical concepts drawn from computer science, sociology, statistics, political science, economics and philosophy. For readers who wish to have a broader understanding of some of their synergies before diving into our design, we begin in the next section of this paper with an extended review of the relevant literature and concepts.  These should be easy for those familiar with any given set of concepts, or those willing to take such frameworks at face value, to skip. Next, in Section \ref{sec:model}, we define the formal objects central to our proposal. We then use these, in Section \ref{sec:mechanism}, to define the proposed mechanism, integrating user experience, algorithmic curation, user content generation and compensation, economic model, etc.  In Section \ref{sec:applications} we apply this model to our three canonical contexts, productivity, news and cultural platforms. In Section \ref{sec:elaborations}, we consider extensions and elaborations of the framework to address issues such as under-served communities and privacy that compose with but are not necessary for the basic exposition of the foundational mechanism.  We conclude in Sections \ref{sec:limitations} and \ref{sec:conclusion} with a discussion of implementation, limitations and directions for future research.

\section{Conceptual and Literature Review}\label{sec:lit}

    In this section, we review core concepts that underlie our design throughout the paper and draw on a diversity of disciplines and historical perspectives with (to our knowledge) limited previous interaction.  We begin by discussing the idea of a social hypergraph, derived from the work of sociologist \textcite{simmel1908}, that is our basic unit of analysis. Second, we anchor and formalize what we understand to be ``common'' in a community using ideas from political philosophy, communications and game theory/economics.  Third, we describe how algorithms from statistics and computer science can harness a variety of input data to detect and classify such communities and thus form an empirical model of the social hypergraph.  Fourth, we elaborate on the principles of the Hutchins Commission (1947) and how these have been and may in the future be translated into algorithmic implementations of community coherence.  Fifth, we discuss the principles of just collective bargaining over data value sometimes referred to as ``Data Dignity''.  Finally, we take a step back to consider the philosophical and ethical approach underlying our focus here.

\subsection{Hypergraphs and intersectional identity}\label{sec:hyper}

    In his 1908 classic \textit{Soziologie}, Georg Simmel describes both the social fabric and the formation of personal identity as the ``intersection of social circles'' (``Die Kreuzung sozialer Kreise'').  In this vision, people and social groups are mirrors of each other.  On the one hand, personal identity is largely constituted by the unique pattern of groups to which the person belongs.  On the other hand, groups are constituted by their members and thus, implicitly, by the diversity of intersecting groups they constitute. \textcite{simmel1890sociale} thus saw the growing feeling of personal distinctiveness (which he calls ``qualitative individuality'') as being an outgrowth of the greater social complexity of modern, urban societies where people went from worshiping, marrying, working, organizing politically, and recreating with the same ``tribe'' of people to doing each with a distinct group.  Unable to manage this complexity, he saw (post) ``Enlightenment states'' as imposing ``egalitarian individualism'' that instead treated people as equivalent members of an overarching nation.  This further exacerbated the challenge for people to manage the complexity of their identity, fueling feelings of isolation \cite{simmel1903}.

    While Simmel is recognized by many as one of the founders of sociology and, partly due to antisemitism his work fell out of the canon as the field professionalized.  When his work was finally translated from German in 1955, Reinhard Bendix rendered it as ``The Web of Group Affiliations'', consistent with the emergence of social network analysis at the time with which Simmel was associated given his previous work on ``triads'' in personal relationships.
    
    It was thus not until the 1970s that Simmel's original vision made its way back into mainstream academic literature.  First, mathematician Claude \textcite{berge} formalized model of social circles as the theory of ``hypergraphs'', extending the theory of social networks to allow many people (``nodes'') to be connected by many groups (``edges'').  Second and independently, sociologist Ron \textcite{breiger1974duality} revived and mathematized Simmel's theory, proposing ``duality between persons and groups'' in statistical sociological analysis.  Within American critical theory, \textcite{crenshaw} emphasized how previous struggles for social equity had neglected the ``intersectional'' nature of identity that Simmel had highlighted, giving birth to the modern movement for ``intersectional social justice''.  Likewise, Bauman saw the problem of social isolation in post or ``liquid'' modernity as arising from a ``model of sociability based on individualism'' where a frenetic need for ``inclusion'' results from the abandonment of an authentic sense of belonging \cite{paleseZygmuntBaumanIndividual2013}.

    Our point of departure in what follows is the Simmelian hypergraphical theory and its elaborations described above.  As such, we are situated in the historical conditions' elucidated by Crenshaw and Bauman while drawing on the formalisms employed by Berge and Breiger to formalize this perspective and thus use it as inspiration for digital systems.  

\subsection{Context and common understanding}
\label{context}

    Social media deal with ideas and imaginaries more than programmatic actions.  As such, a central concept in formalizing the relationship between communities and citizens must be specifying the relationship between personal belief systems and attitudes, and the relevant collective equivalent.  ``Common'' beliefs, attitudes, cultures, etc. have been foundational to the concept of ``community'' throughout its intellectual history \textcite{clarkConceptCommunityReExamination1973, walshRethinkingConceptCommunity1999}, and feature prominently in accounts thereof; a popular set of accounts focusing on this view are those of Yuval Noah Harari, especially \textcite{harari}. 
    
    What precisely ``common'' means here and how it relates to information theoretic constructs of belief was clarified  by philosopher David \textcite{lewis1969} and mathematically formalized by Nobel Laureate game theorist Robert \textcite{aumann1976}. In the Lewis-Aumann account, beliefs that are simply shared are not common or capable of creating shared expectations and behavior patterns typically associated with his expectation of ``community''.  A group of people from around the world may all know English, but until they become aware of the fact that others also know English, they will be stuck communicating in their disparate tongues.  For ideas to be ``common'' in this sense, they must not simply be known by a group of people, but it must be known that all know them, and known that all know that all know them, and so forth.  It is this ``common knowledge'' that allows them to be the basis for common understanding, coordination, communication and action \cite{kuran1991now,malone1994interdisciplinary, clark1996using,chwe1999structure} \footnote{Literal common knowledge is obviously an absurdly stringent condition.  However, as \textcite{rubinstein1989electronic} highlighted, one must exercise care in relaxing it.  Following the standard game theoretic literature, we are referring to probabilistic approximations to common knowledge, such as \textcite{monderer1989approximating}'s $p$-common belief, not deterministic relaxations such as Rubinstein found problematic for coordination.}.

    Events that are simultaneously observed by many thus have much greater power to coordinate collective action than those observed privately, whether the laugh of a boy causing the downfall of a disrobed emperor \cite{chwe2013rational}, a protest causing the collapse of a regime \cite{kuran1991now} or poor economic news causing a bubble to burst \cite{morris1998unique}.  For this reason, common beliefs may differ from the average beliefs of a population if everyone privately changes their minds but has no public forum to express this. While this is typically ascribed to authoritarian regimes, it also occurs in so-called ``open societies'' due to social pressure \cite{kuran1998private} or limits on information flows \cite{morris2002social} — a state of affairs known as ``pluralistic ignorance'' \cite{katz1931students, miller2023century}.  We may thus think of the ``common beliefs of a community'' as partially traceable to (but conceptually distinct from) the (average) beliefs of community members, in addition to being shaped by the affordances of the shared semiotic environment (``media'') they inhabit.

    While most formal analyses consider globally public or private  common belief (with the notable exception of \cite*{chwe2000communication}), sociological analysis of networks (highlighted above) focuses on the intersecting diversity of social circles.  As \textcite{marwick2011,marwick2014} and \textcite{baym2015} highlight, the intersectional nature of digital life requires participants to track, police \cite{nissenbaum2004,jain2023} and code-switch between a diversity of ``contexts'' that have different bases of common belief.  \textcite{wojtowicz2024} formalizes an information-theoretic foundation for this challenge, namely that common knowledge within a context is precisely the basis for effective shared communication, understanding and action.  A primary goal in what follows is to allow a diversity of intersecting communities to achieve this ``contextual confidence'' \textcite{jain2024}, while simultaneously enabling citizens to navigate the resulting complexity.

\subsection{Community detection}
\label{sec:community-detection}

    While ``community-as-common-understanding'' may be a satisfying theoretical construct, it is difficult to observe and measure.  The ``higher-order beliefs'' (beliefs about beliefs, etc.) on which it depends are hard to describe, much less elicit from participants.  Instead, a variety of more easily quantifiable measures have typically been deployed in both research and social media algorithms to ``detect'' and classify their own understanding of what constitutes a ``community'' \textcite{albrechtLivingDifferenceHow2021}.  It is important, in this sense, not to take these models at face value and to remain critical of how they image ``communities'', groups or collectives to constitute, as there are thousands of possible interpretations that may well be operationalised. These may be most easily understood in terms of the input data on which they depend.

    \begin{enumerate}
        \item Graph data: Many early social networks made unweighted ``social graphs'' (sets of bidirectional ``friend'' or unidirectional ``followership'' relationships) the atomic unit of analysis.  As we noted above, this partly arose from the mistranslation of Simmel's ideas in such ``dyadic'' terms.  Whatever the cause, there is a long history of algorithms for community detection within graphs using patterns of strongly connected nodes separated by thinner connections.  These begin with work on detecting non-intersecting or partitional communities, culminating in the celebrated \textcite{girvan2002} algorithm (see also \cite*{newman2004}).  Later work allowed for intersecting communities \cite{palla2005,yang2014,} \cite{xie2013overlapping}, as we discussed above, and applied advances in machine learning \cite{su2022comprehensive} as well as improvements in data to allow for multiple kinds of relationships \cite{kivela2014} to more richly paint this hypergraphical picture.  While there remain important limits to what can be inferred from such dyadic graphs, algorithmic advances have allowed significant inferences about hypergraphical community structure from even such limited data.

        \item Explicit affiliations: In other settings, community affiliation is more explicit.  For example, in some social networks participants subscribe to communities/groups (e.g. message boards, interests groups), affiliate to organizations (e.g. educational institutions or corporations) or disclose/reveal demographic or other community membership valid in context (e.g. linguistic group, citizenship, ethnic memberships).  To the extent they provide a fairly complete account of memberships, they may be used to provide a direct hypergraphical representation.  However, they tend to be incomplete relative to the application (e.g. serving relevant content) and can supplemented with relational/graphical data or additional affiliations can be inferred from the structure of stated affiliations (e.g. a cluster of languages all belong to the same language family or a group of people all identify within the US as ``Asian-American'').  In such contexts, algorithmic methods for community detection and completion in hypergraphs have been developed to enrich the data \cite{tu2018,antelmi2023}.  Again, these are nowhere near a perfect representation of structures of shared beliefs, and not all of the above metrics (e.g., ethnicity) are applicable to all places at all times. But they may more closely approximate them than graphical data alone can, as well as helping make the hypergraph more transparent through intentional affiliations.

        \item Explicit attitudes: Beyond explicit relationships to groups and individuals, attitudes (e.g. ``likes'', reshares, up and down votes etc.) towards content (e.g. text, image or video posts) are a common source of social data.  While such data contain less explicit social structure than do explicit relationships, they more directly express attitudes and beliefs that as we highlighted above are central to understanding common beliefs.  As with graph analysis, the earliest and still most common methods, such as k-means clustering \cite{lloyd1957,lloyd1982}, consider each profile of attitudes to define a point in a high dimensional space and each point belongs to a single community/cluster. \textcite{dunn1973} developed and \textcite{bezdek1984} perfected the most common approach that allows for intersecting communities, Fuzzy C-Means; another canonical approach was developed by \textcite{gustafson1978}. \textcite{aggarwal1999} and \textcite{kailing2004}, among others, proposed approaches that are especially well-adapted to clustering in very high-dimensional spaces, such as those created by attitudes over a diverse range of content, especially when reactions are sparse. A related but distinct tradition decomposes the factors driving attitudes into underlying ``factors'', ``topics'', etc.  These include the classic Principal Components Analysis \cite{pearson1901} and more recent Bayesian approaches such as Latent Dirichlet Allocation \cite{blei2003} and Gaussian Mixture Models.  At their foundation, these obviously have an ``intersectional'' character; if anything, they tend to partition community membership even more smoothly than is sociologically plausible, so that both direct clustering and these latent factor models offer valuable partial perspectives on communities of attitudes, one leaning towards an overly discrete and the other an overly continuous perspective.

        \item Implicit attitudes, behavior and content analysis: While the above techniques involve sophisticated statistical analysis, they all rely significantly on explicit and intentional expressions by participants to guide that analysis.  Other approaches attempt to avoid this, relying directly on the ``natural flow'' of participant consumption to infer relationships, attitudes and interests.  Approaches are broadly like those applied to attitudes, but typically connect these attitudes to direct processing of consumed content using content understanding and classification models \cite{arnab2021,bubeck2023} and often focus directly on prediction of attention and interest, only generating social and community relationships as latent representations rather than explicit targets of interest, an approach often called ``collaborative filtering'' \cite{sarwar2001,koren2009,he2017}.  While such models are often large, complex and lack transparency, they usually also evolve capabilities allowing for some degree of ``introspection'', and thus are often capable of answering natural language or clustering queries that more directly identify some of the models of common understanding and community we discussed above \cite{deldjoo2024}.

        \item Natural language and multimodal expression: A powerful feature of recent models, like those surveyed by \textcite{deldjoo2024}, is that they allow systems to account for multimodal input, as well as providing multimodal outputs.  For example, they can process a variety of responses that participants have to content (e.g. comments in response to content, facial expressions in a recorded video response) and incorporate these into latent understandings encoded in large and opaque ``foundation''/``generative AI'' models \cite{bommasani2022}.  These models can then use this latent representation to generate recommendation and direct answers to questions about provenance, social relationships and social context, such as those around what is commonly accepted/understood in a community , though the tendency of these models to hallucinate and confabulate requires rigorous cross-checking using either other such models or alternative methods.
    \end{enumerate}

    In short, a wide range of techniques have been developed using a broad diversity of data to detect cultural communities/clusters.  More sophisticated and computationally intensive recent techniques open the possibility of using much more ``natural'' data and the chance to approximate more closely the true targets of interest (identifying communities of common understanding), but are less directly transparent/explainable and may be prone to errors.  All can play important and complementary roles, depending on the interaction modality, in identifying and making transparent clusters of common understanding.

\subsection{Bridging and balancing}\label{bridging}
\label{sec:bridging-and-balancing}

    As noted above, the American media landscape prior to WWII was  deeply polarized in its own way \cite{groeling2013media}.  In fact, until the late nineteenth century, most newspapers were explicitly organs of partisan propaganda \cite{gentzkow2006}.  Motivated by mounting academic and social pressure and catalyzed by the US entry into WWII, leading American media magnate Henry Luce supported University of Chicago President Robert Hutchins in 1942 to convene a commission to establish a commission on a ``free and responsible press'', aimed at maintaining a free press in an age of great power conflict by creating a code of press responsibility \cite{blanchard1977}.  While these principles have generally been understood as quasi-voluntary and self-regulatory standards and profoundly shaped professional ethics in journalism, they were also intended as policy recommendations \cite{mcintyre1987} and led to important regulations such as the Fairness Doctrine \cite{pickard2018}.  The recommendations were book-length (1947) but have usually been summarized into five principles of what the press should provide:
    
    \begin{enumerate}
        \item A truthful, comprehensive, and intelligent account of the day's events in a context which gives them meaning.
        \item A forum for the exchange of comment and criticism.
        \item A representative picture of the constituent groups in society.
        \item The presentation and clarification of the goals and values of the society.
        \item Full access to the day's intelligence.
    \end{enumerate}
    
    While these principles also highlight the importance of comprehensive coverage, they can largely be divided into three connected points on which we will focus.  First, there should be a clear distinction between bridging content (facts and shared values reflected in principles 1 and 4) and diverse but sometimes divisive opinion (2-3).  Second, bridging content should form the core of what is presented as ``news'' and should bring the audience together across their divisions.  Third, comment and opinion sections complement this by providing a clear portrait of social diversity and division, ensuring clarity and faithfulness to the range of perspectives. As such, these principles were intended to complement rather than upend constitutional protections for freedom of speech — in particularly against prior restraint and reactions to specific restrictions or viewpoints by public officials \cite{strossen2024free}. We will together refer to the applications of these principles as seeking ``coherence'' of media within a community.
    
    To our knowledge, one of the first attempts to automatically implement coherence in online discourse was \textcite{faridani2010}'s ``Opinion Space'', which, drawing inspiration from deliberative polling \cite{fishkin1991}, arranged a collection of comments based by soliciting participant opinions on the comments and then applying dimensionality reduction to plot them in two dimensions.  In parallel, \textcite{salganik2015}'s ``wikisurveys'' interleaved such real-time feedback with participant generations of new comments, like what occurs in comment platforms like Reddit.  Around the same time, YourView—an Australian forum for policy discussions—incentivized good faith contributions by assigning higher credibility scores to those whose arguments were rated highly by people who disagreed with them \cite{van-gelder2012}.  The Polis platform launched in 2014 and analyzed by \textcite{small2021} combined and built on these earlier attempts by a) adding clustering to the display in Opinion Space to make groups of opinion clearer, b) adding real-time addition of comments and solicitation of opinion as in wikisurveys and c) harnessing a ``group-aware consensus'' metric to surface bridging statements that receive surprising support across existing divides.  This combination of balancing, participation and bridging was particularly influential and Polis went on to large-scale impacts, including facilitating Taiwanese public debate by identifying  compromise on contentious issues like technology regulation and marriage equality \cite{hsiao2018}, inspiring the algorithm behind the Community Notes (formerly Birdwatch) collaborative context provision system on X (former Twitter) \cite{wojcik2022} and sourcing the constitution for a recent large language model at Anthropic \cite{huang2024}.

    There have been other significant innovations and advances in achieving balancing, participation and bridging since its launch. Designs such as that proposed by \textcite{graells-garrido2014} and implemented in services like SmartNews aim to display the positions in more easily interpretable ways, associating them with terms like ``left'' and ``right''.  To merge interpretability benefits while avoiding stereotyping opinions, the Talk to the City project — launched in 2023 by the AI Objectives Institute — processes audio submissions from citizens discussing a public issue by first clustering their responses into topics or arguments, labeling them with natural language, and then providing interactive cluster interpretations, enabling participants to engage in conversations with models representing each opinion cluster. Such ``broad listening'' has spread rapidly, forming an important part of regional and national political campaigns in Japan during 2024\footnote{The term “broad listening” was proposed by researcher Andrew Trask. See \textcite{lichfield} and \textcite{henderson} for a discussion of events in Japanese politics.}.

    The experiments in Japan have also done perhaps the most to extend the user experience of such bridging systems.  During his campaign for Governor of the Tokyo region, candidate Anno Takahiro built a campaign infrastructure that allowed citizens to contribute through a deliberative platform using video, a voice-only call-in line and real-time discussions, including with an avatar of himself.  This allowed Anno to ``crowdsource'' a political platform that received the highest approval of any in the election and allowed him to go from a political unknown to receiving more than 150,000 votes in the course of a month, leading to broader adoption of such systems during the (currently ongoing) national Japanese general election.  Such broader user experiences are likely to be important to systems for generating coherence in the future.

    Still, the ``group-aware consensus'' metric used in Polis, which ranks by the product of support across opinion groups, captures a crucial goal but is, to our knowledge, largely ad hoc. The Community Notes system, by contrast, is grounded in the theory of regression analysis \cite{buterinnotes}, aiming to jointly estimate the position of every rater and note on a single-dimensional political spectrum, and then assess the quality of comments in an “objective” sense by removing the effect of such ``partisanship''. Another approach advocated by \textcite{miller2022beyond} and \textcite{ohlhaver2022decentralized} and deployed in a bridging-based open-source grant-funding system Gitcoin \cite{miller2024case} relies on the \textcite{penrose1946elementary} idea of ``square-root''-based ``degressive proportionality'' for confederal and consociational voting systems \cite{weyl2022}, where more populous subgroups are given sublinear additional weight to account for the correlation in their errors. The growing interest in the use and governance of large foundation models, beginning in early 2023, sparked a wave of new approaches to harness these models for determining and even generating consensual content \cite{fish2023,konya2023,small2023, tessler2024ai}.  Thus far, to our knowledge, all this work has focused on a single audience, spectrum, or conversation — and thus on coherence, not coherent pluralism.

\subsection{Data Dignity}
\label{sec:dignity}

    While our primary focus in what follows is on social dynamics and common understanding, these have implications for the allocation of economic value and incentives in a content ecosystem on which a growing group of journalists, artists, influencers and other creative workers, as well as content moderators and crowd-based norm enforcers, depend.  While it built on much earlier work on imagining a sustainable digital future \cite{nelson1998, nelson1999unfinished, nelson1999unfinished}, data as intellectual property \cite{samuelson1999} and even foundational work on digital privacy \cite{westin1968privacy}, focus on the value and labor on digital platforms emerged with the growth of Web 2.0 platforms in the early 2010s.  \cite{scholz2012,scholz2017b} draws on Marxist theory to highlight how much of the supposed social engagement on digital platforms generates profits for these platforms and thus may be seen as labor.  \textcite{ross2010crowdworkers}, \textcite{irani2015} and \textcite{gray2019} emphasize the precarity and insecurity in which those who do explicit but usually obfuscated work for digital platforms toil, while \textcite{gillespie2018} highlights how central this work is to the moderation and filtering of content online.

    \textcite{lanier3} and \textcite{arrieta-ibarra2018} project this trend forward towards the ``AI future'' imagined by, for example, \textcite{frey2017}, arguing that ensuring fair compensation, general recognition and valuation and decent working conditions for ``data labor'' (a package they have come to refer to as ``data dignity'') will be critical to a sustainable macroeconomy.  \textcite{jones2020} argued that a critical component in achieving this is collective bargaining of data creators through organizations variously called data unions \cite{posner2018}, mediators of individual data \cite{lanier2018a}, data trusts \cite{delacroix2019}, data cooperatives \cite{hardjono2023}, etc.  Such organizations are increasingly forming with support from governments in India and Taiwan.  Seen from this economic perspective, an important role of the communities we emphasize is to serve as intermediaries for data value, both in the direct value generated by the platform and in the growing potential of using content to train foundation models.

\subsection{\texorpdfstring{\pluralism and metamodernity}{⿻ and metamodernity}}
\label{sec:pluralism-and-postmodernity}

    The motivation for our design reflects a growing philosophical focus in many traditions on what we have termed ``coherent pluralism'' \cite{rescher1995} or \pluralism \cite{weyl2024}. This work tends to be grounded in the emerging field of complexity science \cite{waldrop1993,mitchell2009} that extends the older field of cybernetics (Wiener, 1948). Just as the work of internet pioneer JCR Licklider \cite{waldrop2001} and management expert W. Edward Deming \cite{deming2000a,deming2000b,walton1988} did for cybernetics, this research seeks to use complexity science as a foundation for building sociotechnical designs.  Other names commonly used for this include the ``\nameraka'' / ``nameraka'' / ``smooth'' society \cite{suzuki2013}, ``connected society'' \cite{allen2013,allen2016}, ``metamodernism/metamodernity'' \cite{vermeulen2010,freinacht2017}, ``Game B'' \cite{gameb}, and Plurality \cite{Tang2016}.  We are also inspired by related work in consciousness studies that emphasizes the role of ``maps of underlying processes'' (self-awareness) in consciousness \cite{tononi2004} and sees the emergence of collective self-awareness through digital systems as an important step in the progress of cosmic complexity \cite{koch2014,azarian2022}.  Our design can be seen as an attempt to create this kind of ``informational integration'' for a broader social system in a spirit similar to what Deming sought to achieve within a production system.

\section{Model}
\label{sec:model}

    We now describe some of the formal/mathematical ideas we will be using to describe aspects of our proposed design below.  Our goal is not to present a fully mathematically consistent model from which all aspects of the system can be derived (as is common, for example, in economic theory), since the system we envision is too complex to accommodate a formalism that is both sufficiently flexible and not excessively burdensome. However, because we will occasionally use mathematical notation and precisely because of this complexity, we find it useful to clarify some key elements and notation we will use for them.

    Social media platforms are obviously dynamic systems and when relevant we will use the index $t$ as a subscript to denote time. Two primary and one secondary set of participants are central to our analysis. We refer to people who use or potentially use the platform as \textit{citizens}, denoting a generic citizen as $p_i$ and the set of all citizens as $P$. We refer to \textit{communities} within the platform generically as $c_j$ and the set of all communities as $C$. Note that every community is a subset of citizens, so that $\forall j, c_j \subseteq P$ and $C \subseteq 2^P$. We will assume that $C$ is closed under intersections, so that every (non-empty) intersection of a set of communities is also a community; trivial empty communities or those that just include a single citizen we will denote respectively by $\varnothing$ and the citizen's ``name''. For any citizen we denote by $C(p_i)$ the set of all $c_j \in C$ such that $p_i \in c_j$ (the set of all communities of which $p_i$ is a member).

    Finally, we refer \textit{advertisers} on the platform generically as $a_k$ and the set of all advertisers as $A$. Importantly, we will not assume that all citizens, communities or advertisers are necessarily directly observable by the platform or distinguishable from ``fakes''; it may take data for the platform, for example, to determine which of its participants are traceable to an actual person and which communities are ``real'' clusters of interest. At the same time, we also do not assume that there is anything more observable or ontologically fundamental about citizens than communities; both are latent (potentially) unobservable entities.

    We will also refer to the \textit{degree} of membership of citizens within communities. If $p_i \in c_j$, then we will let $s(p_i; c_j) \in (0,1)$ by the relative standing in which $p_i$ is held within $c_j$, relative to other members $i \in c_j$; we do not specify here precisely how this normalization works. Similarly, we denote by $d(c_j; p_i)$ the \textit{devotion} of $p_i$ to $c_j$ relative to other communities in $M(p_i)$, symbolizing how relatively important her membership of $c_j$ is compared to her membership of other communities.

    A central analysis unit will be content. The set of all content on the platform we denote by $M$, with typical element $m_i$. Such content may belong to topics $T$ with typical element $t_m$, which bear the same structural relationship to content that communities bear to citizens (viz.\ a piece of content may belong to several topics, which are closed by intersection, etc.).

    We will sometimes speak of the belief, adherence or affect of citizens or communities in a piece of content; for short we will refer to this as ``belief'' but this is obviously more relevant for some contexts than others. For example, for an entertaining piece of content the relevant interpretation will typically not be belief but rather some mixture of enjoyment and comprehension, or interest and curiosity. For citizens this is intended to correspond roughly to the usual decision-theoretic notion of subjective probability or in some cases to a preference/utility parameter; to focus on the first such interpretation and without loss to the second we assume these attitudes are always between $0$ and $1$. In the case of communities, we interpret this parameter as a measure of common belief or higher-order belief about adherence. In both cases we denote belief by $b(m_i;p_i)$ or $b(m_i;c_j)$. We see belief having two sources: attitude, representing a latent tendency to believe in the content, and exposure, which is the degree (or probability) that the citizens or community has encountered the content. We denote the former by $a(m_i;p_i)$ or $a(m_i;c_j)$ and the later by $e(m_i;p_i)$ or $e(m_i;c_j)$.

    Some assumptions are crucial to our analysis and we clarify them here.

    \begin{enumerate}

    \item \textit{Belief generation:} For citizens, we will typically assume that $b(m_i;p_i)=a(m_i;p_i)e(m_i;p_i)$; that is, belief can be the product of attitude and exposure\footnote{Something roughly similar is true for communities, but given other assumptions below, this relationship cannot directly hold for both citizens and communities.  There is likely a superior model to either of these that hold symmetrically across the two; this is an interesting direction for future formalization of the ideas here.}.  One obvious implication of this is that exposing citizens to things they have a good attitude towards is more ``productive'' in generating belief than exposing them to things they have a poor attitude towards.

    \item \textit{Personal and community beliefs:} We assume several things about the relationship of citizen and community belief of which cannot be fully and concisely mathematically formalized, but which are worth being clear about nonetheless\footnote{Beyond the notation involved, fully formalizing the model in a logically consistent way would likely involve adding exogenous drivers of changes in beliefs and making beliefs and attitudes equilibrium objects, a significant distraction given our goals here.}.

    \begin{enumerate}
        \item \textit{Monotonicity:} We assume that community belief is strictly increasing in the belief of community members, holding the belief of other members fixed. Similarly, we assume that citizen attitudes are strictly increasing in the belief held by all communities she is a part of, holding fixed the beliefs of other communities she is a part of and other factors driving attitudes.
        
        \item \textit{Linearity under consensus:} If all members of a community have the same degree of belief in a piece of content, then the community can tend to have this belief.
        
        \item \textit{Standing:} The greater a citizen's standing in a community, the larger impact an increase in her belief has on the community belief. Similarly, the greater devotion of a citizen to a community, the greater is its impact on her belief.
        
        \item \textit{The role of consensus:} \textbf{Perhaps the most important assumption we make is that consensus can be critical in generating belief.} When beliefs are heterogeneous in a community, the community's common belief will tend to be lower than the average (even weighted by standing) belief\footnote{Mathematically, this property is known as Schur-concavity.}. This property captures the idea that when iterating beliefs, whomever believes the least acts as a ``weakest link'' dragging others down as it interrupts the chain of ``knowing, knowing that others know, etc.''. This tends to imply, intuitively, that common beliefs are harder to sustain in diverse communities. We apply the same principle to citizens: heterogeneous beliefs in content among communities they belong to make it hard for them to sustain a positive attitude towards content. This again reflects the Simmelian idea that internal division drives a sense of alienation and lack of integrity. On the other hand, we acknowledge just as well that diverse beliefs may be productive to yet more beliefs, or may be the motive of productive compromises. 

        \item \textit{The role of structure:} Beyond the above and even harder to formalize, we assume that this premium on consensus depends not just on the profile of direct participants, but on the whole hypergraph structure, in a particular way.  \textbf{Division among prominent subcommunities makes common belief hard to sustain.}  Thus, if among the most important and commonly divided subcommunities all have similar common beliefs in a content, the overall community belief in it will be like these even if, for example, there is significant disagreement within these subcommunities or between these subcommunities and other subcommunities.  The idea here is that beliefs spread and organization occurs socially, so that divisions among significant subcommunities are more likely to lead to dissensus and conflict than are simple fluctuations in personal beliefs or among small subcommunities.  This mirrors the widespread concern over polarization, the emphasis in the Simmelian peace studies literature on the value of cross-cutting cleavages \cite{coser1956functions,lipset, horowitz1985ethnic}) and the extensive literature on innovation in groups on graphs \cite{granovetter1973strength, milroy1992social}.  We assume a similar dynamic applies to citizens, with the most important inhibitors of their belief formation being divisions between their most central community attachments.  We refer to the small (2-7) principal subcommunities of a community by $\sigma(c_j)$.
        
    \end{enumerate}

    \item \textit{Scarcity and allocation of attention:} We assume that exposure is scarce, in the sense that there is some limit on $\sum_i e(m_i;p_i)$ for each citizen $p_i$ and that the platform can, by its policies, reallocate this exposure among content for the citizen.
    
    \item \textit{Value of coherence:} Finally, while we have not introduced much about money or monetary transactions above, when discussing economic issues below we make several critical and closely related assumptions.
    \begin{enumerate}
        \item \textit{Communities value coherence:} We assume that communities value ``coherence'' in the sense of giving exposure (labeled as such) among their members to content likely (conditional on exposure) to generate common belief and to a balanced set of content across leading subgroups (content likely to generate common belief within leading divisions of that community). Our logic directly derives here from the Hutchins Commission.
        
        \item \textit{Citizens value (internal) coherence/integrity:} We assume the same principle applies to citizens, driven by the Simmelian idea that internal incoherence or lack of understanding of one's internally conflicting instincts has psychic costs. This idea is also prominent in a variety of psychological (e.g. Freud and Jung) and religious (e.g. Taoism, Buddhism, Hinduism) traditions.
        
        \item \textit{Advertisers value common belief:} We assume that advertisers value not just citizens believing their content separately but also communities believing their content, because some important consumption decisions are collective/social in nature \cite{veblen, galbraith1958, baudrillard, ewen, miller1997material}.
    \end{enumerate}

    We further assume that at least many of the citizens, communities and advertisers have some financial resources that they are willing and able to use to help achieve these values.
    
    \end{enumerate}

    With these premises in place, we propose a design for the allocation of exposure adapted to them, aimed at achieving the coherence desired by citizens and communities and taking a share of this as profit for the platform.

    \begin{figure}
        \centering
        \includegraphics[width=\linewidth]{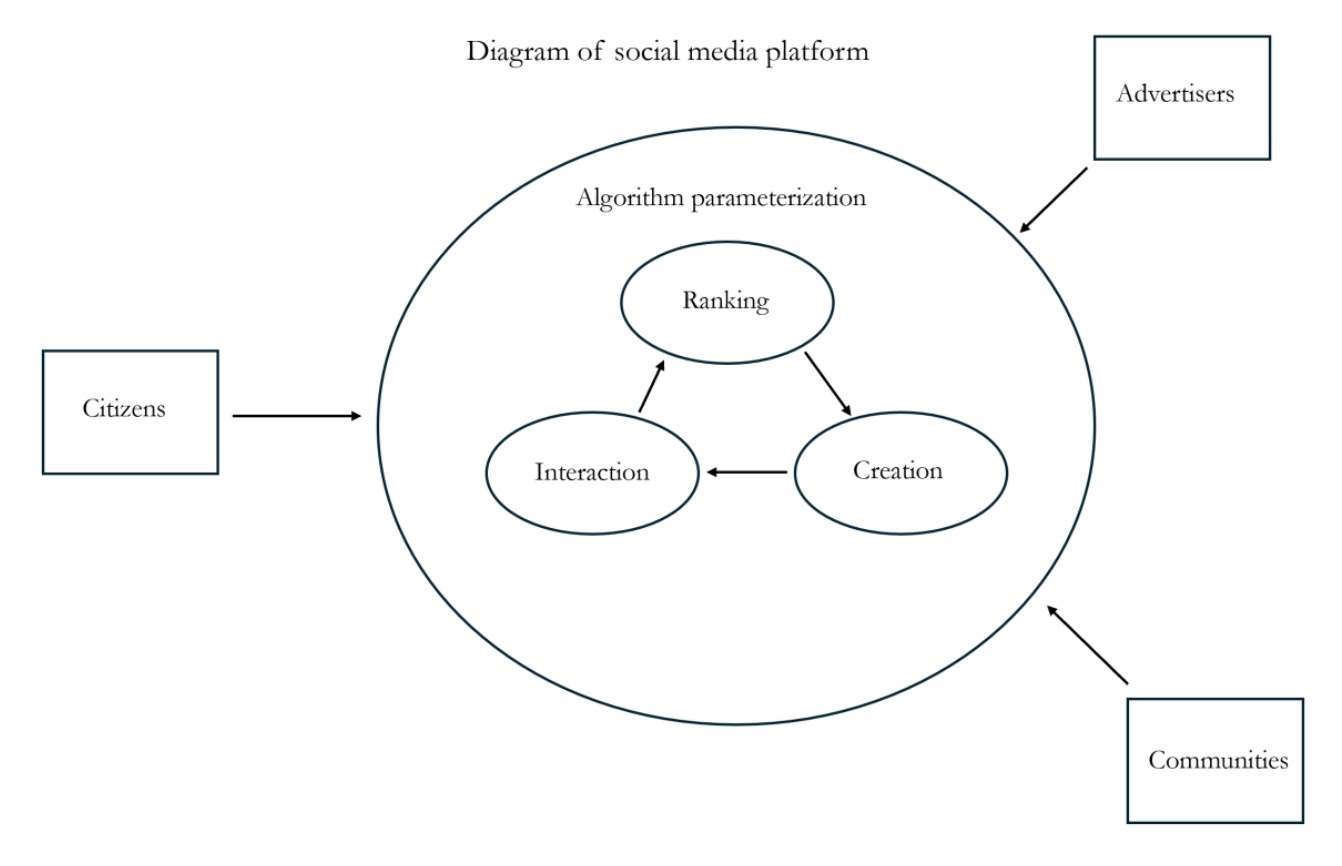}
        \caption{Diagram of the social media platform with participants, enrollment, algorithmic parametrization and loop of actions.}
        \label{fig:loop}
    \end{figure}

\section{Mechanism}
\label{sec:mechanism}

    Because a social media platform is an on-going system involving a diverse range of asynchronous and partially interdependent actions by a variety of actors, it is not amenable to description in a linear sequence.  Instead, we describe the system we imagine in terms of a core \textit{loop} of activities, a process of \textit{parametrization} of an algorithm that conditions how this loop functions (and thus changes over longer time scales) and a process of \textit{enrollment} into the system of actors.  Each of these pieces must be understood through several \textit{lenses}, describing parallel perspective on them.  Figure \ref{fig:loop} above illustrates and summarizes this structure.

\subsection{Loop}

    As we noted above, the central idea of social as opposed to traditional media is that ``peers'' simultaneously and in parallel \textit{interact} with, \textit{create} and implicitly by their actions and through an algorithm \textit{rank} the content each other see.  This contrasts with a traditional media environment where the roles of audience (interaction), staff (creation) and editors (ranking) are clearly distinguished and separated in distinct people.  In social media the flow from staff to editors to audience is this replaced by a circular relationship or loop in which content created by citizens is exposed to (other) citizens, who interact with it and in the process generate data used to rank this content for still other citizens, which in turn helps content creating citizens understand what content is of interest, shaping their future creation of content and so on.  Below we specify more precisely the process we imagine playing out in each of these functions.

\subsection{Parametrization}

    Our focus is on algorithmically mediated content curation, as we see this as core to defining social media.  In such curation, signals from citizen interactions with content and possibly machine analysis of the content directly is used to determine the prioritization and thus reach of the content in the presentation of content to other citizens, such as in a ``news feed'' or ``for you feed''.  Some features of such an algorithm are fundamental and structural, such as what citizen signals it considers, metrics it aims to maximize or the architecture of the models it uses to connect signals to goals.  Other features, which we will refer to as ``parameters'', are adjusted or tuned more frequently.  We consider a critical aspect of the system conditioning behavior inside the loop to be the less-frequent-but-still-ongoing adjustment of parameters in response to actions taken by citizens, communities and advertisers.  In more traditional social media setting, a classic example of this would be advertisers placing advertisements targeted at some group, leading those with various interaction characteristics to see that advertisement (more frequently); another would be (on some platforms) citizens choosing to subscribe or follow a feed, leading to an increased probability (in interaction with other signals) of that citizen seeing content from that feed.  We imagine similar but more sophisticated interactions between model parametrization and participant actions that we detail below.

\subsection{Enrollment}

    The system we describe has three important distinguished roles of participants, though presumably some persons might in practice play several roles at once: citizens, communities and advertisers.  To participate, each needs to be formalized within the system and the rights to control their role secured.  In a typical social media environment, this will occur through enrollment, account creation and some kind of initializing activities (e.g. expressing interest in a set of topics) in two separate systems for citizens and advertisers; how communities are created varies across settings.  In what follows, we outline how enrollment might occur and the way we imagine communities being added to the system.

\subsection{Lenses}

    Social media simultaneously are (at a minimum) technical, social/psychological and economic systems: they are defined by algorithms for computing content display based on data, settings for the joint interaction with and creation of content by citizens and systems of allocation of economic resources and generation of financial surplus.  These three aspects are deeply intertwined and jointly causal, but operate in very different conceptual models, typically studied using different tools by different disciplines.  Therefore, for each of the above elements of the system, we describe how we imagine each of these three aspects of the system working and connecting to the others in what follows.

\subsection{Specifications}

    We now provide a detailed specification of the system components above, seen from these various lenses.

\subsubsection{Interaction}
\label{sec:interaction}

    The central element of any social media system is the interaction between citizens and content.  Content is displayed (with a reach determined by the ranking process as discussed below) and citizens consume it and provide real-time feedback.  We now describe those interactions from a technical, social and economic perspective in the system we imagine.

\paragraph{Technical}

    The technical nature of the interaction between citizens and content depends heavily on the type of content and thus will differ across platform type as we discuss extensively in Section \ref{sec:applications} below.  However, there are two key systematic changes relative to the status quo we imagine making to the nature of the interaction: displaying social provenance and attending to critical responses.

    First, we imagine every piece of content displayed to a citizen $p_i$ to be marked with at least one sign of \textit{social provenance}.  This indicates the community or communities the citizen is a member of where this content is bridging or balancing.  The precise user interface in which this is displayed will again vary across format, but it is possible to describe generically and concisely the general semantics.  For bridging content, the message would be that in community $c_j$, where $p_i \in c_j$, this content is widely viewed and accepted.  One natural form of display for such information would be an indication as common in many existing sites that ``trending among Americans'', but likely with slightly adjusted wording for clarity such as ``Americans are into'' or ``Americans agree on'', depending on context.  For balancing content, the message would instead be that in community $c_j$ this content is divisive, representing one side of a debate.  A natural short message here would be ``Americans are debating'', with the ability to follow a link from the word ``debating'' to another screen where the side of the debate and leading content on sides of this debate can be navigated.  There, an LLM could summarize the main arguments made by a community about a given topic, their underlying premises, and how they compare to another community's. In some settings even these short verbal messages are likely to long and instead something more visual would be a better fit, such as a picture representing the community (e.g. an American flag) and a color or smaller superimposed icon representing consensus (e.g. green, or a handshake) or dissensus (e.g. red, or fisticuffs).

        \begin{figure}
        \centering
        \includegraphics[width=\textwidth]{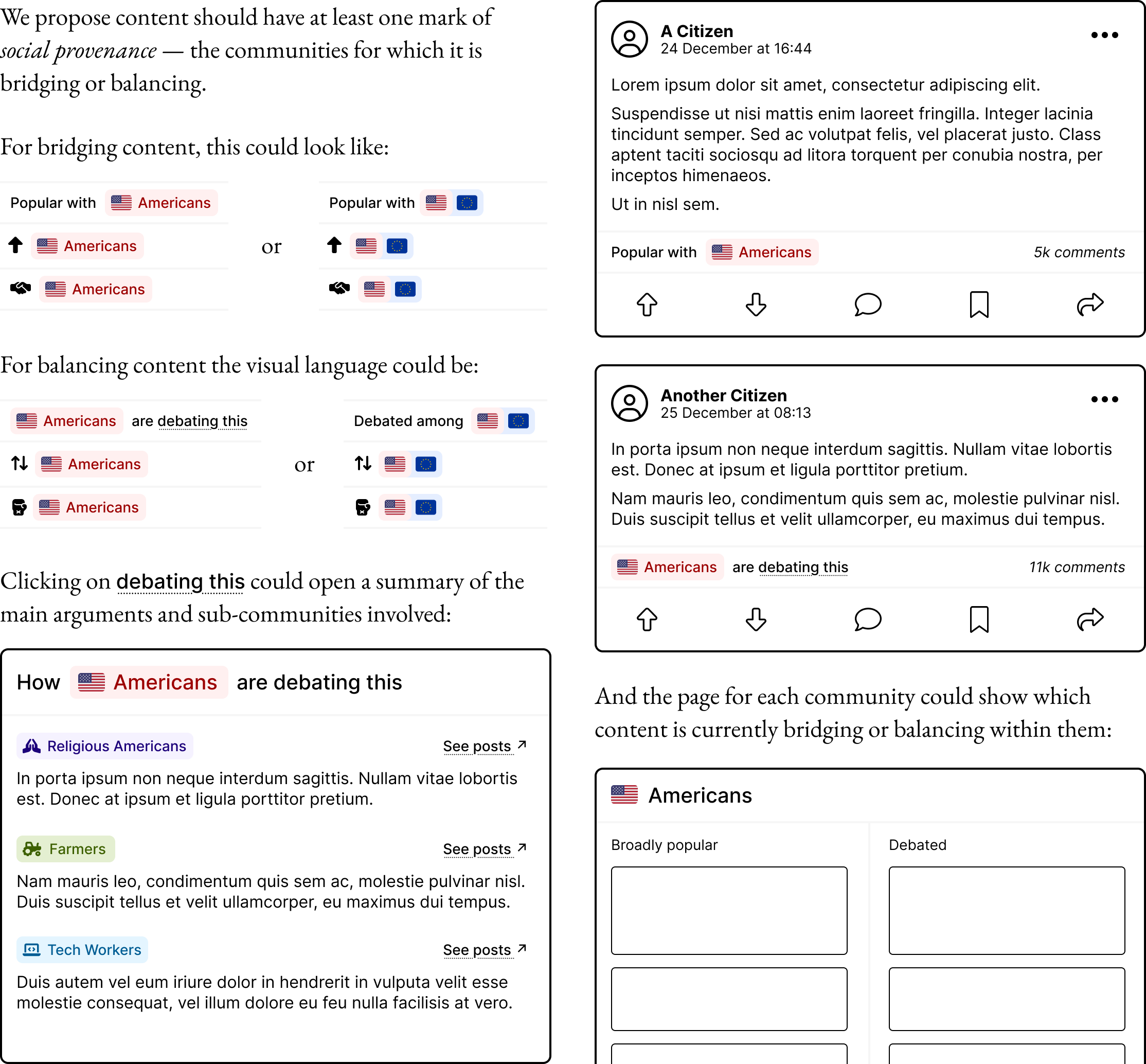}
        \caption{Illustration of how social provenance and context could be made transparent.}
        \label{fig:social-provenance}
    \end{figure}

\begin{figure}[t]
    \centering
    \includegraphics[width=\linewidth]{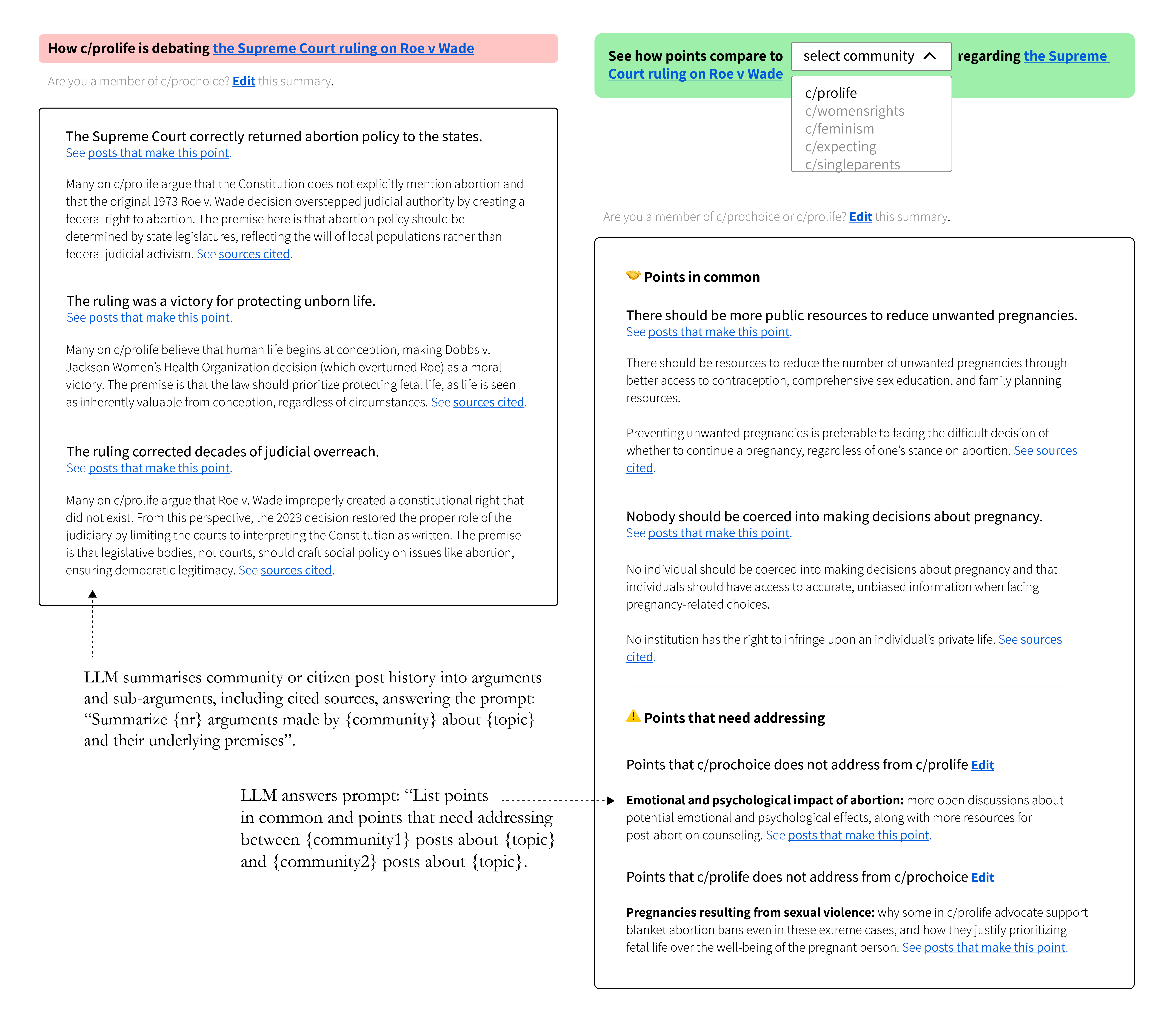}
    \caption{Illustration of how an LLM could be used to summarize the points a community makes about a given topic, and how these compare to adjacent communities}
    \label{fig:enter-label}
\end{figure}

    In many cases (depending on parameters below), a piece of content might serve different roles in different communities.  For example, an American Republican might see a piece of content labeled as bridging within the Republican community they belong to but divisive in the US overall.  Furthermore, given the intersectional nature of identity in our model, these potentially conflicting social reactions are important to citizens and would be labeled as such.  If content were bridging across the communities the citizen is part of, they might, for example, be labeled with that citizen's image and the green color or handshake; if the content was divisive ``within that citizen'' (viz. viewed positive in some of her communities and negatively in others) it would be displayed with her image and the conflict indicator, inviting the citizen to explore the texture of that debate.

    Second, we imagine the system actively soliciting or detecting negative/critical responses to content, as well as the positive responses that are already core to most social media interactions.  Again, the format of such input will vary by setting, but to provide the data required for the ranking protocols we imagine below, negative affect is an important input.  In some contexts, this can be implemented simply by adding a thumbs down reaction, and some social media systems include these.  In other contexts, with more implicit feedback, it is likely best inferred from analysis of comment reactions or from training models to detect patterns of viewing or scanning indicating outrage or offense.

\paragraph{Social}

    The social/psychological intent of the changes above is to create in participating citizens a sense of common interaction and thus common understanding.  Citizens would be led to, correctly, believe that whenever a piece of content appears prominently in their feed as bridging within a community that this is simultaneously being viewed by others in that community are also seeing this content and seeing with the same bridging content.  Conversely, whenever they see content appearing prominently as being contentious they would be led to believe, correctly, that this is contentious across persistent lines of division within this community, is being portrayed to other citizens in that community as such and that contrasting perspectives on similar topics or themes are easily accessible to other citizens in the community through the exploration feature described above.

    To further foster this sense of connection, it would likely generally be advantageous for the platform to create complementary features that enable citizens to voluntarily make more private connections with other members of the communities.  The platform might offer adjacent opt-in features that allow people to discover community members, possibly who are also members of other intersecting communities with them.  For example, a citizen heavily viewing Sikh and Boston area consensual content might opt-in to connect with others at that intersection, helping them find a local Sikh temple, study group or friend to explore their shared faith. Another example would be people who are active in their local community and are also amateur gymnasts, who might form a private chat group in which to discuss how best to improve their local gymnastics facilities.

\paragraph{Economic}

    Interactions between citizens and content will generally trigger micropayments based on the principle of ``pay per impression''.  The direction of the payments will depend on the type of content and other parameters specified below, but the general types and directions of flows can be described more broadly.  Funds will generally flow from \textit{sponsors} to \textit{creators}.  The former include advertisers, communities supporting content that aid coherence within them and citizens subscribing to receive content that aids their coherence.  The latter includes the platform itself as a source of its revenue to fund operations and those who create the content (as described in our next point) that fuels participation and thus underlies the revenue of the platform.  While the notional principle of micropayments would underlie the flow of funds within the system, the implementation would likely in practice bundle such payments over time before actual transfers are made and possibly involve various constraints, averages or caps to address the unpredictability of prices or payments they could otherwise create.

\subsubsection{Creation}

    All content in the system is assumed to be uploaded by citizens or advertisers.  Creators seed the content into the system with either social stake (viz. quantified, community-specific reputation) within some community (which they earn as we discuss below) or, in the case of advertisers, an economic arrangement with the community.  Depending on the platform type, the content will have to conform to technical and possibly legal standards to be displayed, as analyzed by likely at least partially automated content screening systems. Because this is the most straightforward step, we do not break it down by lenses here, except to note that in our design creators should be given a clear view of not just the generic reach of their content, but its social situation (the communities where it plays different role and the prominence it gains through those roles).

\subsubsection{Ranking}

    The most important changes we propose are to the way that content is prioritized for citizens, which we now describe. Algorithmic curation of content is central to how we define social media. Even the simplest and most transparent or ``democratic'' ways of organizing content, such as ordering them by positive reactions or by the number of other citizens one citizen follows who have reposted them, are algorithms that take the behavior of other citizens as input to determining the prominence of content for a particular citizen.  More sophisticated and typically  opaque algorithms have proven more adept at executing on often problematic goals, such as maximizing user engagement.  However, such more sophisticated algorithms are also critical to even defining more appealing objectives, such as coherent pluralism.  At the same time, those goals depend on a degree of ``transparency'' or self-awareness of the social system, to which the black box nature of current ranking systems is a potential challenge.  We aim to define a ranking system that incorporates elements of these more sophisticated systems but, by tying them to elements of the interaction experience above, to harness this greater sophistication to elevate the collective self-understanding of citizens and communities.

\paragraph{Technical}

    We imagine each piece of content, $m_i$, being assigned (using content analysis) a set of topics or themes $T(m_i)$ and, for each community $c_j$, using the interaction data from Section \ref{sec:interaction} above, a measure of the degree of interest from that community, $\iota(m_i;c_j)$\footnote{ We imagine this interest being dynamic and time-dependent in a way we do not emphasize here, to accommodate the notion that, for example, more recent content should be of greater interest.}. We also imagine each piece of content being classified, using the methods from Section \ref{sec:community-detection}-\ref{sec:bridging-and-balancing}, as either \textit{bridging} as quantified by a bridging score $\beta(m_i;c_j)$ or as divisive and characteristic of a strict subset of the principle subcommunities of $c_j$, $\sigma(c_j)$, with a degree of association $\delta(m_i;c_j)$. Every divisive piece of content would, where possible, be associated with counterparts with similar scores $\delta$ but associated with non-intersecting subset of $\sigma(c_j)$. For a divisive content $\ddot{m}_i$, we refer to this affiliated set of content as its \textit{balancing} content as $\kappa(\ddot{m}_i;c_j)$\footnote{While in some cases it may be desirable to define bridging and balancing content \textit{by topic}, we expect communities to generally be interested in bridges within a community across persistent, cross-topic divides that may lead to persistent fractures and to value the crossing of these fractures, even if they lead to other persistent, topic-specific divisions.}. Each piece of content then receives an overall \textit{community score} that depends positively and interactively on its interest and the greater of its bridging or balancing strength; this might be more general, but for the moment we will simply denote it $\psi(m_i;c_j) \equiv \iota(m_i;c_j)\max\{\beta(m_i;c_j),\delta(m_i;c_j)\}$. Each community $c_j$ also has an associated parameter, $\lambda(c_j)$ that is set as described in the next subsubsection that represents the weight placed on the interests of that community by the algorithm.
    
    We similarly imagine applying the same techniques to define such scores for citizens: $\iota(\cdot;p_i)$, $\beta(\cdot;p_i)$, $\sigma(p_i)$, $\delta(\cdot;p_i)$, $\kappa(\cdot;p_i)$, $\psi(\cdot;p_i)$ and $\lambda(p_i)$\footnote{ Note that while we define $t$ at the citizen level for completeness and consistency, it generally will not differ significantly from the additive community-based analog and thus does not play a significant role.}. These should be defined by the same analyses for communities, taking the communities a citizen is part of as analogous to the subcommunities of a community. Principal subcommunities then represent leading lines of fracture within the citizen's identity. For example, a citizen may be part of a religious conservative group but progressive political groups; she may be Jewish but in a relationship with an Arab; they may an American of Chinese origin with warm feelings towards China. As such, it makes sense to think of bridging and balancing within a citizen as well.
    
    We consider the ``feed'' of each citizen to be a platform/algorithm-determined allocation of attention paid by each citizen to each piece of content $e(\cdot;p_i)$. The precise quantitative interpretation of this is complex and context-specific (e.g. does it index where on a feed content is displayed or the expected action of the citizen given this display pattern?), but a variety of metrics such as display location, size/prominence, time spent, etc. should be comonotonic with it.
    $$e(\hat{m};p_i) \equiv \frac{\lambda(p_i)\psi(\hat{m};p_i) + \sum_{c\in C(p_i)} d(c;p_i)\lambda(c)\psi(\hat{m};c)}{\sum_{m\in M} \left[\lambda(p_i)\psi(m;p_i) + \sum_{c\in C(p_i)} d(c;p_i)\lambda(c)\psi(m;c)\right]}$$

    In other words, we imagine ranking taking place based on the coherence score of the content for the citizen and the communities she is a member of, all weighted by the $\lambda$ parameter and the communities additionally weighted by the citizen's devotion to that community. This overall score is compared to that of all other content to determine the weight on the content.

\paragraph{Social}

    Following this ranking algorithm and the interaction format described above, citizens would see their feed constituted by content that is bridging or balancing, as well as drawing attention, in communities they are members of and for them personally.  They would have the opportunity to adjust their devotion to each community manually, as well as likely have it adapt to their behavior depending on the amount they chose to engage with content for that community.  In addition to ranking, the interface should allow citizens to navigate content by community, to screen out other combined influences on their feed.  Because equally devoted citizens in a community will see content ranked with similar prominence, this helps confirm the impression conveyed above of a collective consumption experience and is a key reason why certain explicit features of the ranking function are important to preserve to create common knowledge.

\paragraph{Economic}

    While we do not anticipate any direct exchange of financial value based on rankings alone (instead, as noted above, we anticipate this to happen based on interactions), we anticipate changes to formalized social capital accompanying ranking. When a piece of content $m_i$ created by citizen $p_i$ achieves a high score in community $c_j$ (a high $\psi(m_i;c_j)$), the standing of that citizen in that community $s(p_i;c_j)$ should rise to reflect the credibility and relevance of their messages in that community. Conversely, when a citizen first posts a new piece of content in that community, she should be able to expend some of this standing to give it initial prominence (a high effective initial $\psi(m_i;c_j)$ score) that allows it to gain the initial exposure that can allow interactions to provide the data for ranking the content. This community-specific reputation acts as a ``community currency'' (Prewitt, 2022) to allow those with standing to get a hearing for their content. We discuss broader tools for exploration of new content and for initialization into community standing below. Advertisers may also be sold such standing by community representatives as a means of ``artificially'' boosting their content, as we describe below.

\subsubsection{Parametrization}

    A crucial floating parameter set above is $\lambda$, which defines the weights placed on communities and citizens in defining feeds and ranking.  We now describe the origin and role of this parameter.

\paragraph{Technical}

    We imagine $\lambda$ being adjustable in continuous time, but likely at a lower frequency than daily interactions, by the associated citizen or community.  These adjustments would come with effective associated costs, however, which we discuss shortly.

\paragraph{Social}

    Citizens can raise their $\lambda$ value to increase the ``personalization'' of their content, where personalization here specifically means the serving of content that strengthens their sense of personal integrity and understanding of the tension they face in the facets of their identity.  Communities can raise their $\lambda$ value to increase the prominence of their cohering content thereby strengthening their self-understanding.  Because attention is fixed, there is a zero-sum nature to this competition, but at the same time all aspects provide contextualizing perspective to strengthen identity.

\paragraph{Economic}

    While we do not imagine changes to $\lambda$ having any \textit{direct} financial implications, they influence which content is served to whom and who is the sponsor versus the recipient of content.  As citizens raise their $\lambda$ values, they will begin paying for more content as subscribers and the higher communities set $\lambda$ the more impressions they will end up paying for.  Either communities or citizens can choose to boost their $\lambda$ value at no additional expected cost by accepting, depending on which, either personalized or community-targeted advertisements that will run in the same context as that unit.  We anticipate, given the greater capacity of community administrators compared to citizens, for citizens to exercise less direct agency in shaping this parameter and to expect greater clarity on the financial implications when making an adjustment, while communities are likely to more directly select and filter appropriate advertisements.

\subsubsection{Community}

    Communities are the most distinctively central actor in the design we propose and we imagine their formally joining the system more indirectly than other actors, through a combination of emergent identification and fostering, in conjunction with active registration and enrollment.

\paragraph{Technical}

    For a community to become active in the system, two hard conditions and one soft condition need to be jointly satisfied.  First, the community must be detected using techniques as described in Section \ref{sec:community-detection} as relevant to the overall social hypergraph.  Second, an administrator judged as a legitimate representative of the community by the platform must step forward and administer the community account; we will discuss how legitimacy may be determined or an administrator encouraged to step forward shortly.  Finally, a softer requirement is a sufficient group of the proposed community members detected algorithmically choose to opt into and remain in the community, given that their participation is consensual as we discuss below.  If this latter condition fails, the community may continue to exist formally but will play little role.

\paragraph{Social}

    To enroll communities, the platform effectively needs to match a behavioral/algorithmically-identified community with a legal or natural person who can represent this community.  Doing so will require at minimum algorithmic interpretation of the participation to enable platform administrators to understand who to approach.  In some cases, this is likely to be sufficient and the associated organization unambiguous.  For example, for geographic groupings corresponding roughly to the areas of legally recognized administrative divisions (e.g. a city or a nation state), the natural authority would be the office of the corresponding government with responsibility for communications and public media.  In other cases, there may be multiple potential representatives, such as a cryptocurrency ecosystem with no official representative organization but many prominent community leaders.  Such examples will require careful judgment by the platform using its internal metrics as well as having to potential rely on direct democratic input from citizens, as we discuss shortly.  Finally, some cases, especially in emerging or under-organized communities, are likely to initially have no natural representative, such as the community around a decentralized emerging protest movement like \#MeToo.  In such a case the identified community could remain in a register of communities waiting to be ``claimed'', searchable by candidate community leaders who wish to initiate a registration.  In such cases it may also be helpful for  the platform to actively work to foster the building of organizational capacity in the emerging social group, as we discuss in Section \ref{sec:underserved-parties} below.

    Most communities in the system are unlikely to have existing direct democratic structures: while a few may be democratic nation states and others democratically governed civil organizations (like some unions and religious organizations), most will likely have other governance structures (e.g. for-profit corporations).  On the one hand, the system should respect this diversity and not attempt to force communities to conform to some imposed ideal of governance.  On the other hand, given the algorithmic derivation of communities and the ``volunteer'' basis on which they might be claimed, with some frequency there might be mismatch between what community members collectively see as their legitimate representatives and the current administrator of a community.  A reasonable compromise would be to allow community-initiated changes to community management based on sufficient vote (potentially weighted by participation), subject to applicable trademark laws in relevant jurisdictions, but to set the threshold for such a change at a level that deters transient complaints or pressure campaigns.  This might offer an interesting design space for experimenting with innovations in voting in digital spaces \cite{weyl2024}.  However, given that community management is largely a source of cost (viz. to support raising $\lambda$) and that the goals of the community are largely given by the algorithmic structure described above and implicitly democratically determined by community members, it may even be possible for many organization to participate in community management, contributing jointly towards the financial support of a high $\lambda$ and negotiating jointly with advertisers.

\paragraph{Economic}

    As noted above, communities act as sponsors, depending on the level of $\lambda$ they set.  Registration will thus generally require the same financial verification, deposits, means of payment, etc. platforms typically require from advertisers.

\subsubsection{Citizen}

    Citizen enrollment in the platform is much closer to standard enrollment in a social media platform, with a few adjustments to the structure we propose.

\paragraph{Technical}

    While enrollment should generally follow best practices from analogous social media platforms, expression of interest in at least one community as either a participant or a creator would be necessary to begin receiving or contributing content that receives attention.  Furthermore, to receive some initialized standing within a community that could be used to ensure posts receive attention, a citizen would want to demonstrate that they are a validated and/or valued community member.  One simple way to do this generically across a variety of communities could be a personhood credential as analyzed by \textcite{adler2024personhood}.  However, many communities are likely to have their own community-specific forms of identity verification and stake establishment.  For example, a national community might request some kind of national identity card, while a club might request proof of club membership.  Furthermore, the importance of formal affiliation with communities might in turn open the possibility of using such memberships as a basis for identity custody and recovery through social recovery schemes as described by \textcite{ohlhaver2022decentralized}.  In addition to community enrollments at account creation (as is familiar from prompts to follow people or topics when first setting up a social media account at present), citizens would be periodically queried for their consent to join or remain in communities that emerge or to which their participation is detected, as discussed in the previous subsubsection.

\paragraph{Social}

    Enrollment in the system, while generally aligned with existing social media environments, carries connotations, expectations and verification requirements that are somewhat different from those expected in most current social media environments. The closest analog is likely the current Fediverse environment and especially the extensions of it proposed by Christine Lemmer-Webber and her colleagues.  Our environment assumes more strongly than many social media systems that the citizen has (at least within communities) a single account and that each account corresponds to a real person.  However, such ``unique identity'' may be divided among several accounts with separate communities or, as we discuss in Section \ref{sec:privacy}, be partially occluded in one community from others based on where that citizen is participating at this moment.  This aims to maintain and defend the importance of ``contextual integrity'' to community common knowledge \cite{nissenbaum2004, nissenbaumbook,nissenbaum2011contextual, jain2024}. This approach is in the spirit of ``selective disclosure'' or ``meronymity'', where partial characteristics are selectively disclosed, but sufficient to avoid manipulation of community norms and democratic participation in a contextually-specific way \cite{immorlica2019verifying, ohlhaver2022decentralized}.

\paragraph{Economic}

    While it would likely be desirable to allow enrollment without registration of payment information, there would be more pay-for features (based on boosting $\lambda$) in our environment than in some fully advertising-supported social media environments and there might be the potential for citizens to receive payment on net from communities and advertisers for engagement if they choose not to invest this in a higher $\lambda$ and thus more net sponsorship.

\subsubsection{Advertiser}

    Even more than with citizens, advertiser enrollment should follow roughly the pattern of existing social media platforms, especially given that the idea of connection to and targeting of social groups is a well-established mechanism of such systems.  We therefore only briefly address translational issues to our set-up.

\paragraph{Technical}

    Advertisers would enroll as in a standard social media platform.  However, they have the chance to target not only citizens who have opted into personalized advertising as discussed above, but also communities.  While their relationship to citizens would be like the present setting on the advertiser's side (differences on the citizen's side were described above), the relationship to communities would resemble the banner advertising environment, with more direct negotiation and selection by community administrators of appropriate content.  Communities might also use a standing system and community member signals as with content creators to improve evaluation of when advertising is appropriate for their community's shared context.  While direct citizen engagement in selecting advertisements they desire is unrealistically burdensome, citizens could choose to apply similar metrics in terms of coherence to filter the advertisements they receive around their identity.

\paragraph{Social}

    Advertisements would remain a significant component of the system we describe.  However, the social context of these advertisements would change significantly.  Rather than being primarily targeted at individuals, they would appear in community contexts with the intention in many cases generating common understanding, ``buzz'' and action around the advertised product, with the agreement of community leaders.  In some cases, advertisements deliberately seeking to position themselves relative to controversies might align with divisive rather than bridging content and be labeled as such within communities. In either case, advertisements would thus be a commercial component of the generation of common understanding.

\paragraph{Economic}

    While the platform would retain the some of the advances that have fueled the greater return-on-investment features of targeted advertising (such as quantifying viewership and paying based on impressions), other metrics would be inappropriate and need to be replaced as a basis of payment in this environment.  For example, ``pay-per-click'' \cite{kapoor2016pay} and ``pay-per-conversion'' \cite{jordan2011multiple} models will typically misfit an approach to advertising focused less on direct inducement of actions than on creation of social meaning.  Joint impressions, based on social relationship, and creation of common understanding, on the other hand, might finally be closer to quantifiable in this model.  Just as a science of measurement of advertising grew up around online advertisements, if this model succeeds a similar but more sociologically grounded marketing science would be critical to the success of the systems we imagine.

\section{Discussion}
\label{sec:discussion}

    We do not formalize behavior or the mechanism sufficiently here to formally demonstrate any properties of the mechanism.  However, we believe it is important to describe properties we believe are likely to result from it that motivate the design and why we believe these are likely to hold both so they can be measured in applications and experiments, as well as guiding further adaptation of the mechanism and attempts at formalization in the future.  We now list, describe and briefly argue why these properties are likely to hold, focusing on outcomes that apply across a range of contexts; in the next section we turn to features relevant to specific categories of platform application.

    \begin{enumerate}

    \item Social self-consciousness: A central goal of our design is ``transparency'', though in a somewhat different sense than the term is typically used in the computer science literature, which typically focuses on simplifying models to make them more easily explainable \cite{rudin2019stop}.  We aim to create greater alignment between the underlying complexity of the social system in which citizens are embedded, as modeled imperfectly but more richly than possible in the past by contemporary digital systems, and the understanding of the citizens of that system.  We do this by projecting the complexity of existing models onto a somewhat simplified hypergraphical framework that can be meaningfully conveyed to citizens as described above.  While this model is still simplistic compared to the full richness of a deep network representation, the fact that it can be communicated and given real feedback by the social groups it models, rather than turning these richer models towards extremely simplistic ends such as engagement.  This bridges the gap between the richness of the models and the simplicity of the human feedback and goals given to them, helping to create social systemic self-awareness.

    \item Depolarization and the social fabric: Recent research increasingly emphasizes several factors in creating escalating affective and toxic polarization that our design addresses.  First, one of the most effective interventions in reducing extreme polarization has been found to be reducing higher-order misperception\cite{iyengar2019origins}.  Correcting such misperceptions is central to the design of the system we propose, as it is engineered to provide common knowledge as primary output.  Second, the physical sorting of people by the most salient identity categories correlated to primary political polarization is an important driver of polarization, while drawing out more complex cross-cutting cleavages is a critical contributor to depolarization \cite{mason2016cross,klein2020}.  The emphasis in our design on multidimensional, intersecting and cross-cutting identities, and the ability to help citizens manage and achieve integrity in face of these, is directly designed to address this driver.  Finally, by allowing communities to directly pay for bridging content, the system directly makes depolarization a product.

    \item Connection and community: Many have remarked on the difference between early internet communities (such as Usenet) and some contemporary social media (like Reddit, Facebook groups and Clubhouse) on the one hand and the more dominant contemporary ``personalized feed'' experience of social media on the other hand.  While the former foster the critical conditions \textcite{baym2015} identifies for building community online, the latter tends to undermine these and thus is arguably and important source of the feelings of isolation that many highly-online young people increasingly experience \cite{haidt2024}.  A primary design feature of the system we propose is to create, in an integrated, multi-community, feed-like interface, the indicators of common values and beliefs, as well as mutual awareness and ability to deepen conversation, that enable the creation of communities in Baym's spirit.  The hope would be that it therefore leads to more of the feelings of connection and community fostered by the former type of digital community, while offering some of the diversity of interaction of the latter.

    \item Collective action: Most ambitious human activities depend on coordinated/collective action, because of economies of scale or more broadly ``supermodularity'' \cite{supermodular}. In politics, protest movements and the passing of legislation both require large scale coordinated efforts. In business, coordination across many workers and business units is the essence of both management and strategy.  In culture, everything from festivals to memes depend on many people acting in at least parallel if not tightly coordinated ways.  The central condition for collective action is common understanding of the action, roles and responsibilities, timing, and common interests.  Achieving such common knowledge is the centerpiece of our design.  A deeper formal analysis of the knowledge structures it creates compared to other social media environments would be very valuable.

    \item Social dynamism: A primary promise of the internet was to allow the emergence, and primary vision of the philosophies that inspired it \cite{dewey1927}, was to allow the emergence, self-awareness and organization of previously disparate social groups, often themselves brought into interdependence by emerging technologies.  While in many ways the internet has gone a long way towards accomplishing that goal, allowing marginalized communities to meet and form community across distance \cite{gray2009out}, especially in recent incarnations it has not empowered these emergent communities to coordinate and organize themselves.  Our system creates both the capabilities necessary to achieve this and, by charging communities, an economic incentive for platforms to support the development of formal organization by such emerging communities.  Together these can help foster the social dynamism that Dewey saw as the essence of democracy and thereby helped inspire the development of the internet \cite{weyl2024}.

    \end{enumerate}

\section{Applications}
\label{sec:applications}

    While the generic design and potential benefits described above apply across settings, there are at least three quite distinct types of social media platforms where the details of operation, most important benefits, and user experience are likely to differ radically.  In this section we discuss such differences and the patterns we expect in these three cases.

    \begin{table}[p]
        \centering
        \includegraphics[width=\linewidth]{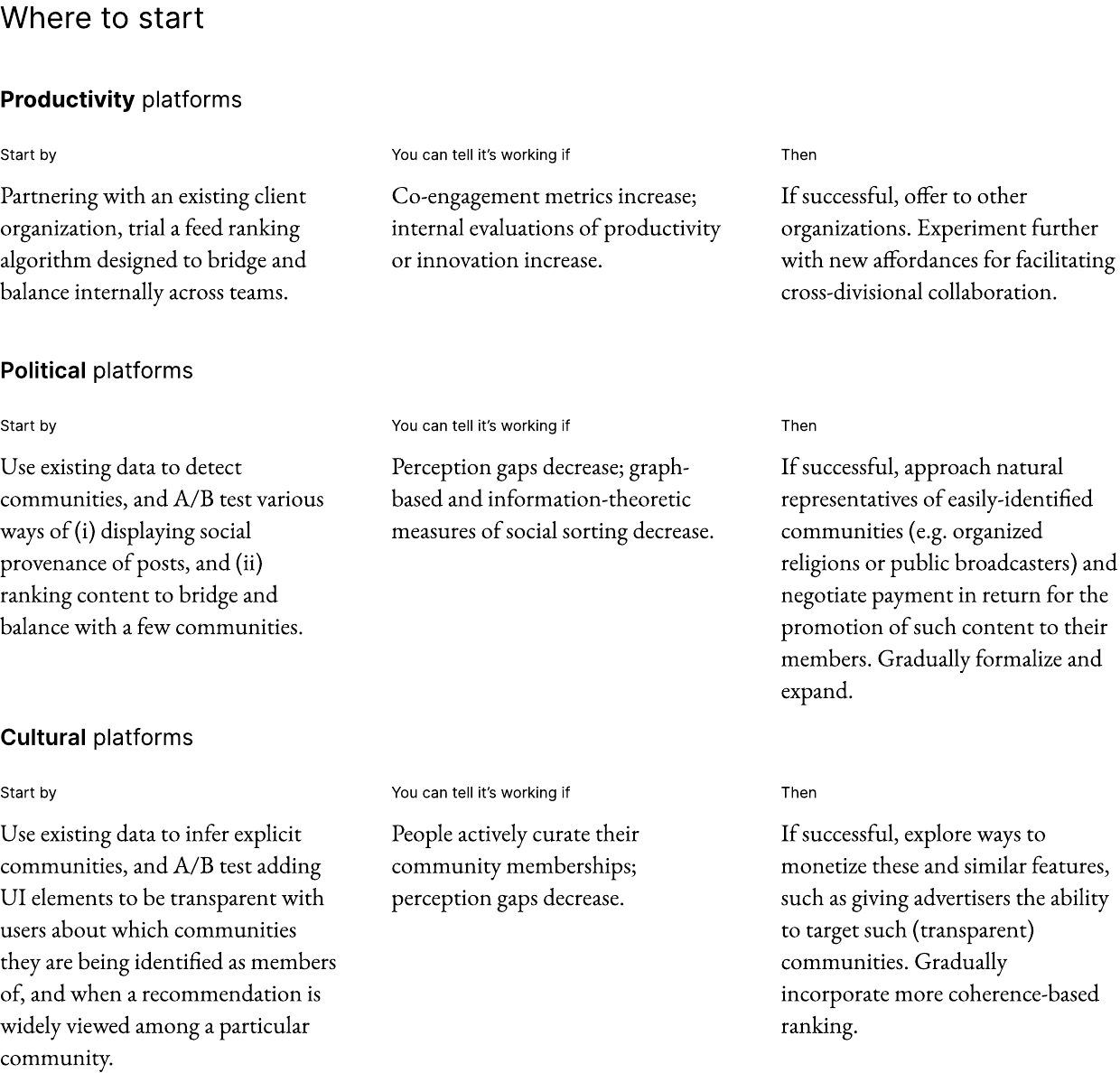}
        \vspace{1mm}
        \caption{Concrete suggestions for how to start experimenting with community-centered social media, and what metrics to look at to tell if it's working.}
        \label{fig:transition}
    \end{table}

\subsection{Productivity}

    While most social media are supported by advertisements and thus are driven to focus on engagement, this pattern is not universal.  A smaller but still prominent segment of the market is platforms in the productivity space that derive their primary revenue from businesses in relation to their current and potential employees.  While there are a variety of such platforms, those that are both most prominent and most closely fit this description are several owned by Microsoft, employer of one of the authors of this paper: LinkedIn, GitHub and to a lesser extent Teams.

    Several features of such platforms align more closely with our design than the other categories of platforms we consider below.  First, these platforms typically involve explicit and usually verified affiliation of members to a variety of organizations, including employers both past and present, credentialing organizations (e.g. universities or coding academies) and projects.  While not the only source in such a setting of communities, it does make their identification simpler, more consensual, and easier to identify representatives.  Second, these platforms derive their primary revenue from organizations already, though not from precisely the model we imagine.  This would make it easier to begin spinning up a new business model along the lines we suggest.  Finally, the platforms ``brand promise'' or image is less around entertainment and enjoyment and more around productivity, making the potential collective productivity impacts of the system we describe clearly aligned with the platform mission.

    What, though, is the precise ``pitch'' for the productivity benefits of our system?  Much of the discussion above focused on social and political reasons to favor \pluralism.  However, the economic case is at least as compelling and often easier to measure.  One of the most persistent themes in the management literature is the way that traditional, hierarchically branching management structures create internal, cross-divisional conflict within organizations impeding broader goals \cite{lawrence1967,walton1969,galbraith1973}.  A particularly fitting illustration is that under CEO Satya Nadella, Microsoft (mentioned above) has tried to overcome its extreme reputation for such internal conflict by promoting a culture of ``One Microsoft'' work.  Perhaps the area of greatest attention to this challenge surround innovation and ``intrapreneurship'', where the structure of an organization conflicts with an emerging business necessity that cuts across current divisions \cite{christensen2013}.  In response, many leading firms have created processes specifically geared to overcoming these challenges of ``disruption'' and much of the (significant fraction of) work time devoted to cross-group meetings focuses on achieving understanding and alignment across various cross-divisional boundaries. In fact, the primary revenue source for GitHub is solutions designed to allow cross-divisional collaboration on proprietary but internally open software.

    A natural application of our proposed system to this goal would be to serve employees of a company posts from other employees that help them discover, on important business and technological issues, internal divisions and points of consensus across that difference.  This could happen within the business overall, but also within organizations and cross-divisional, topical working groups, implying the intersectional structure of our model.  Some of these features would be most appropriate for a setting such as Teams or Slack with a strong privacy/security model, as many key topics of alignment involve confidential business information.  However, these environments typically have a more limited emphasis on intersecting, algorithmically curated feeds.  Those would fit better with an environment like LinkedIn or, to a lesser extent, GitHub.  Likely the ideal setting would be some mixture of these, where community provenance is used to track both consensus and privacy/security issues, perhaps as an outgrowth of collaboration between products like LinkedIn and Teams.

    In some contexts, the emphasis on privacy and security could go even farther; for example, a leading role of alliances like the North Atlantic Treaty Organization is to achieve alignment and interoperability among member militaries.  While there are to our knowledge not currently widely used social media applications with sufficient security properties to facilitate the sharing of such extremely sensitive information, one can imagine that were they to be developed, some of the features of the system we discuss above might be helpful in achieving cross-divisional or international, intra-alliance alignment and understanding.  We return to questions around privacy and security in the next section.

    Some other important productivity applications, however, would not intersect nearly as much with issues around confidentiality.  For example, recent literature on the development of technology hubs like Silicon Valley and Boston's Route 128 corridor emphasize the central role of universities and venture capital networks in providing ``neutral'' spaces in which people working at different companies can interact and explore potential new ventures, allowing for regional dynamism \cite{saxenian1996}.  While many have, with limited success \cite{lerner2009}, attempted to imitate these experiences and build regional innovation hubs, few have systematically fostered the linkages across organizations that this work suggests is critical to such ecosystems.  With regional communities as customers, the system we describe could offer a platform for understanding and bridging the divides between organizations within an industry-region.

    To begin implementing our proposal, a natural step would be to work with preexisting organizations with which a productivity platform has a commercial relationship through its current business model and explore payments for the curation of their employees' feeds.  Employees could be asked to consent to this (partial) curation in exchange for some benefit, or simply as a test.  Metrics of co-engagement with content could be shown to companies and tracked over life of the trial, being compared against internal evaluations of productivity on content relevant projects, potentially being rolled out on a division-by-division basis to offer a point for comparison.  The intervention could then be offered to other organizations and scaled to intersecting affiliations to curate a full feed.

\subsection{Political news}

    A second category of social media most closely fits the motivations throughout this paper: those that are used for primarily for discussing news, politics and current affairs.  The most prominent among these is X (formerly Twitter), along with its primary competitors: Meta's Threads, BlueSky and the ``Fediverse'' platforms based on the W3C Activity Pub standard such as Mastodon and Truth Social\footnote{An interesting contrasting competitor here is Reddit, which shares much more of the explicit community ethos we discuss, but largely without the algorithmic curation we take as an assumed starting point. In that sense, our design is something of a bridge between the ``X-like'' environment and the Reddit environment and we could almost as easily discuss how to potentially introduce more automated curation into the Reddit environment while maintaining its ethos. We leave that an exercise to readers in the interest of brevity}.

    All these platforms share important features that fit much of the motivation of our proposal, though there are significant differences among them, especially between the Fediverse platforms and the rest, that would affect implementation as we return to below. First, all share a primary orientation to text, with some use of images and even more limited use of video.  Second, all focus on a ``microblogging'' format with generally short posts, with responses, reposts and ``likes'' or signs of approval as the primary active feedback.  Third, connections are primarily based on ``follows'' in a directed person-based social graph, where citizens choose to subscribe to or follow other citizens. Finally, monetization is primarily based on advertisements and/or subscriptions by citizens.

    The motivation for our design is perhaps the clearest and most aligned with our discussion throughout in this setting of the three we discuss.  However, there is one reason for out design that we did not fully articulate above because it is specific to this context.  A primary aspiration in the early days of platforms like Twitter is that they would empower the formation of self-governing political movements and thereby increase citizen's sense of civic empowerment and self-efficacy \cite{papacharissi2015affective}.  Typically, however, as documented by \cite{tufekci2017}, while these movements have been effective in rapid organization, gathering of attention and expression of sentiment, they have struggled to make collective decisions and achieve a coherence and consensus of demands that allows them to achieve enduring broader public sympathy and lasting political change.  This has led to significant backlash, disillusionment and decline in support for democracy in much of the world \cite{mounk2018}.

    Yet, as noted above, this disillusionment has not been universal. In Taiwan, the different social media institutions we described above led their Sunflower movement of 2014 towards a coherent set of demands that helped bring in a new government in 2016 that achieved broad public approval and empowered citizens, especially the youth, to build consensus for and pass dozens of pieces of citizen-sourced legislation \cite{huang2021towards}. At least partly as a result, the 2022 wave of the International Civic and Citizenship Education Study found Taiwanese youth had the highest sense of civic self-efficacy in the world \cite{iea2022}.  The design of our system imitates, combines, extends and scales aspects of the Taiwanese ecosystem is thus motivated significantly by the aspiration to see such empowerment made available to a broader diversity of communities participating in such platforms.

    A natural first step to implement our system on a centralized platform like Threads or X or an open one like BlueSky would be to use existing data to identify communities and bridging and balancing content within those communities.  Experiments could then be run with highlighting such content within those communities and marking it in the way we describe.  Reactions to and effects of such experiments could then be analyzed and presented to representative of easily-identified communities within the hypergraph, such as organized religions or public broadcasting authorities.  Negotiations could then begin for payment for the promotion of such content as a pilot, with citizens being offered discounted or additional services for opting into such an option.  This could spread from a single community outwards, with additional features that depend on such interactions being gradually incorporated.  Implementation on Fediverse platforms would likely be significantly more complicated, as it might need to involve citizens joining multiple instances and these instances cooperating, possibly using a protocol like Spritely on which some of the developers of ActivityPub are now working.  We will not discuss this case further here in the interests of brevity, but it is an important area for future research.

\subsection{Culture}

    The final and most widely utilized category of social platform is those that focus on cultural and entertainment experiences, such as TikTok, Instagram, YouTube and, more ambiguously, Facebook.  Also in this category belong ``metaverse'' platforms with rich virtual reality content, such as Roblox and Decentraland.  While these platforms have garnered great attention for their intersection with news and politics, their primary focus is elsewhere, which is likely an important part of why they have roughly at least an order of magnitude more participants and attention than the other categories of platforms.

    Like the other categories, these platforms share several key features that differentiate them from the other categories.  First, their focus is on content offering a sensorily richer experience than text, such as images, video, audio and immersive experiences.  Second, while they share some of the explicit feedback features of the other platforms, data on ``direct reactions'' (length of engagement, sequences of engagement, attention, scrolling, movement etc.) tends to be a more central data source for the platform to determine content recommendations.  Third, monetization is overwhelmingly, though not exclusively, based on advertising and compensation of content creators is a much more important part of platform economics.  Whether relationships and social affiliations are explicit and direct or implicit and indirect differs across examples.

    Because these platforms are not primarily economic or political, the reasons for and benefits of designing for \pluralism on them should not be primarily political or economic, though we strongly believe that the foundation of political and economic flourishing is the robust social and cultural fabric such a platform can provide.  Instead, the goal must be to strengthen that fabric itself.  Our design aims to promote an intersectional popular culture, constituted of a diversity of collectively self-conscious and overlapping cultural communities.  Broadcast media and mass market products have long been long been critiqued for homogenizing and sterilizing culture, while marginalizing alternative influences and sources of creativity \cite{adorno1947,galbraith1958}.  The move, manifest in both the ``counterculture'' of the 1960s and the movements of the 1970s and 1980s to bring faith back into public life, to overcome this ``high modernism'', have succeeded in fragmenting culture, but, many believe, at the cost of a sense of coherence of identity and sense of belonging \cite{jameson1984,harvey1989,rorty1998}.

    The natural ``solution'' to this conundrum as many have remarked lies in the well-trodden vision of \textcite{tocqueville1835}, \textcite{arendt1958} and \textcite{putnam2000} where a diversity of intersecting social communities at various scales constitute both ``training grounds'' for the building of broader culture and fabric in a ``quilt'' which ties together a societies as sub-hypergraphs.  Achieving this requires simultaneous diversity of groups, to avoid polarization, and the popular self-consciousness that constitutes popular culture.  It is precisely this combination that our system aims to promote.  We aspire that by building this diverse, intersectional popular culture, citizens will feel empowered to embrace, navigate and manage the diverse facets of their identity, feeling belonging in a range of social groups that are partially in tension.

    Given the diversity and size of these platforms, we focus on only one example of a potential implementation pathway, relevant to TikTok and its primary competitor Instagram Reels.  Given the implicit nature of most social affiliations and inferences in this environment, the first and most natural step would be to begin identifying explicit social groupings and simply highlight to citizens the communities they are being identified as members of and the trending of videos in which is primarily driving their recommendations.  Such transparency, even before any coherence-oriented content organization, would be an easy way to test the interest in and receptivity of communities to labeling of content with social provenance.  If citizens are receptive, it would then be natural to look for ways to monetize this feature, making community participation and then sorting content for coherence natural pathways, as discussed in other applications above.  Advertising into these communities seems particularly appealing, given the long-standing connections between popular cultural meaning and successful advertising campaigns we have highlighted.  Experiments with such campaigns and their cultural impact would be a natural next step, helping strengthen the commercial case for our model in this context.  The banning of TikTok in several jurisdictions and potential leadership transitions in the process have created a premium on the kinds of transprency our design woudl instantiate, making it a relevant moment for such experimentation.

\section{Elaborations}
\label{sec:elaborations}

    While encompassing many elements, the core of our design leaves out a range of important issues that are not necessary to understand the basic dynamics.  For brevity, we cannot cover these possible elaborations in even the detail we gave above for the basic mechanism, but we wish to highlight and some directions they might be developed in the future and sketch how a solution might be made consistent with our broader structure.

\subsection{Underserved parties}
\label{sec:underserved-parties}

    A central source of revenue for our imagined system is payment for coherence by communities. Such payments are plausible for communities with organizations, resources, coherent representation and the ability to provide public goods like coherence.  However, a core motivation for the internet, dating to the work of \textcite{dewey1927} and repeated frequently in internet culture by thinkers like \textcite{ugarte2009} and \textcite{srinivasan2022}, is that many important emerging social groups and communities have no history of self-awareness much less organizing institutions.  A crucial potential of our system as noted above is precisely to empower the emergence and development of such communities online.  Yet it is in their nature to be initially unlikely paying customers.

    On the one hand, this is a very common problem in the development of platforms.  For example, most operating systems begin with a pool of potential developers with experience on adjacent operating, but very few current applications.  A major part of platform strategy is providing tools, support and encouragement to develop an ecosystem of applications that make the operating system attractive to consumers and profitable for the operating system creator.  The work of empowering the development of self-governing online communities that can help support the platform is analogous.

    On the other hand, creating a business imperative for the platform to support the development of such communities offers an important novel opportunity to fund the development of common infrastructure many such communities could rely on to develop, such as tools for collective decision-making, funding of institutions, support for forming legal organizations to represent them, safeguarding of collective funds etc.  While there has been growth in recent years of movements around online communities, there is limited direct funds available to support the infrastructure of self-governance for such communities \cite{scholz2017a, schneider2021, horowitz2021}.  By making such communities a critical commercial counterparty, our system could offer a funding stream to support critical infrastructure to actualize the Deweyian dream of empowering emergent communities.

\subsection{Privacy}
\label{sec:privacy}

    Most of our analysis above assumes, as is true for some platforms (especially most political/news platforms today) that all relevant data for making inferences about content are public.  While this may be a reasonable approximation in some context, even there many would aspire to move beyond this model of extreme transparency, with all the risks it holds for ``context collapse'' \cite{boydcomplicated} and erosion of personal identity \cite{zuboff2019}.  In other contexts, such as most productivity and cultural platforms, this poses a completely unacceptable threat to commercial confidentiality and/or personal privacy.  Furthermore, there are key features of our system that would seem to lend themselves to enriching the potential for thoughtful ``privacy''.  As \textcite{nissenbaum2004,nissenbaumbook,nissenbaum2011contextual} famously emphasized, privacy is typically more about contextual duties of confidentiality to a community than about individual rights.  Because of the central role community explicitly plays in our system, it may offer the possibility of formally defining more appropriate notions of ``contextual confidence'' than is possible in most present digital environments \cite{jain2023,jain2024}.  While this is an enormously rich topic well beyond the scope of this paper, we will sketch some possible directions.

    There are two distinct frameworks in which to think about privacy in this setting: one relying on the platforms as a trusted third party and another truly based on decentralization, as would be most appropriate in, for example, a federated system.  Because we return more extensively to issues around federation in the next subsection, we will focus on the trusted third party context and only briefly discuss how this might be extended to a decentralized setting.

    A crucial feature to note about our proposed system is that each created piece of content is posted into one or more community contexts, requiring staking of reputation within that context.  This provides a natural marker to limit the spread of that content outside the intended community context.  There is always a challenge of ensuring that community members do not ``reshare'' to other contexts. However, there are a number of platform features that could discourage or undermine this threat including disabling direct cross-community resharing, using cryptographic tools like designated verifier signatures to make it impossible for community members to prove content was actually created by a particular person to those outside that community (especially effective given the rise of deep fakes) and communities reducing the standing of members who engage in inappropriate resharing.

    Beyond content itself, there may also be concerns about the way the system uses interaction, feedback and affiliation data to identify social groups.  This could likely be addressed in several ways.  First, much of the clustering and analysis could be performed by the platform, providing only aggregate outcomes and not personally attributable affiliations or feedback for each member.  Second, affiliations and feedback within a particular community are the critical data, so that even if these were made available to a community, much of which would be appropriate, would not require sharing other behavior of that member and members could choose to occlude some identity features from communities where it might be inappropriate.  Third, to the extent citizens selectively suppress identity affiliations this will not significantly bias estimates of the metrics of interest unless it affects the divides and consensus across social groups.  For all these reasons, most of the benefits of the system could probably be maintained while preserving most privacy benefits.  Furthermore, the community structure could allow  representatives or even votes among communities to determine sharing policies or decisions.  For example, if content were shared and popular within a community, with the consent of the creator the community could determine whether it was appropriate to share on as a representation of that community to an intersecting or broader community.

    Implementing the same features in a fully federated and community-sovereign way would have additional complications and perhaps limit some functionality, but should generally be feasible.  Again, communities need only share aggregates of reactions to shared content with other communities to provide them signals needed to compute bridging and balancing scores.  They would typically (though not always) tend to have an interest in doing so, to ensure their community was fairly represented in other settings.  This would likely lead to stricter privacy and somewhat reduced functionality, with some additional technical challenges, but lead to fairly similar results.

\subsection{Portability and federation}

    A primary aspiration of many technology reform and alternative social media movements recently has been greater ``portability'' or ``ownership'' of the social graph by those that are part of it \cite{mccourt2024}.  A central challenge in implementing such portability is that relationships have no natural single ``owner'': they belong collectively to those participating in the relationship \cite{immorlica2019verifying}.  Exit by one member of a community, even if they can bring with them a record of all their bilateral relationships, does not help them reestablish a community elsewhere unless other community members simultaneously move with them.

    A more practical way to establish credible and relevant portability may therefore be to ensure that significant communities (clustered subgraphs) are collectively self-aware and have community control over their relationships and data in a way that allows for them to exit collectively.  Doing so would not ensure that every citizen in that community would see all their communities migrate; they might still have communities that stay on the platform.  However, at least they would, for an important subset of their uses, know that a significant community to them will migrate with them, making the new platform more relevant to them.  Whether facilitating such portability is in the interest of a private platform or more likely a feature that emerges from an open standards process or government regulation is a distinct question, but the community-based model suggests a more feasible way to structure such portability than many currently-imagined models.

    Of course, if such portability were to become a reality, the social media platform landscape would likely become significantly more fragmented.  This need not necessarily be a significant problem if data federation, as discussed above, allowed for exchange of key features of social sentiment to allow the identification of cohering content across communities.  This would, however, require more substantive data sharing standards than have to our knowledge been embedded in existing Fediverse systems, making this an important topic for the development of future standards, especially given that the specific relevant data is likely to differ between e.g., productivity, political news and cultural platforms.

\subsection{Model training and content value}

    While the primary role of content and community data in the system above is to facilitate its functioning directly within the platform environment, data from social media are believed to be an increasingly important component of training large foundation models, though precise details of training data are often confidential.  This trend seems likely to continue and perhaps even amplify in coming years, making the question of how to treat platform data in relation to model training a prominent one, with platforms like Reddit and X increasingly taking a proprietary approach to data generated by citizens.

    One prominent reaction to this has been the movement, especially around blockchain communities \cite{dixon2024,mccourt2024} for ``personal data ownership''.  Yet as some pioneers of the movement for citizen economic rights to data, such as Jaron \textcite{lanier3}, have argued, individuals rights to data face many challenges to actualization given the complexity and lack of bargaining power individuals have \cite{lanier2018a}.  Exclusive control by centralized and typically asymmetrically powerful platforms, however, carries similar risk to those by model developers and in some cases (X, Meta) they are actually one and the same already.  Many have therefore argued for data intermediaries, often called ``unions'' \cite{posner2018}, ``trusts'' \cite{delacroix2019}, ``cooperatives'' \cite{hardjono2019data} or ``coalitions'' \cite{tang2021}, to bargain collectively for data value.

    Just as with data portability, a natural role for communities in this setting is capture and appropriate scale at which such intermediaries could be formed.  Communities could act as fiduciaries for their data for economic and governance as well as social and direct platform purposes, negotiating terms on which data in that community could be used and receiving data payments to defray community costs, reduce advertising, maintain infrastructure and compensate the creators of valuable data as highlighted in Section \ref{sec:mechanism} above.  Because each citizen would belong to several communities, this would essentially implement an important part of \textcite{lanier3} and \textcite{lanier2018a}'s vision of people being part of several contextual intermediaries, each bargaining for a relevant cluster of data, and because of the clustering would also address many of the concerns of \textcite{acemoglu2022too} and \textcite{bergemann2022economics} around the way correlated data degrades citizen bargaining power.  The platform could construct higher level infrastructure, whether direct model training or direct negotiations with model creators, to facilitate these intermediaries agency in exchange for a share of surplus, while leaving to the communities the most important substantive privacy, use, federation and economic questions.

\subsection{Accountability}

    Every platform moderates content to some extent to address gross violations of the law, such as non-consensual intimate material, commercial fraud, incitements to violence, etc. While on centralized platforms this is typically taken on by the central administration, in federated systems each instance takes responsibility for policing content, with a hybrid system prevailing in settings like Reddit that delegate significant authority to administrators of subreddits.  Recently, some scholars such as \cite{section230} have suggested augmenting liability for clear violations with softer liability for algorithmic amplification of problematic content.  A variety of other compliance and regulatory constraints exist in various global jurisdictions.

    An interesting possibility opened by the community structure we suggest, building on the Reddit model of governance, is to delegate in a subsidiary manner \cite{hasinoff2022} governance to communities wherever possible.  The platform overall could enforce accountability for communities to legal standards in domains of operation, giving communities the choice of whether to comply or accept that their community will be inoperative for citizens in relevant jurisdictions to avoid platform liability. It could also pass onto communities liability for violations, but allow each community to determine the standards applied.  Furthermore, the platform could create community compliance tools allowing, for example, communities to monitor cross posting for potential challenges to the community-specific rules and norms.  Such subsidiarity would mimic some of the division-of-powers properties of federal and confederal governance systems and provide partial localization of some of the most controversial issues in platform governance \cite{gillespie2018}, without entirely leaving subscale communities to manage all regulatory burdens alone.  From the perspective of regulators, thoughtful approaches would likely attempt to attribute responsibility to different actors in the system in accordance with their roles: to the platform for features arising from the overall architecture of the algorithm, to communities for the setting of parameters and policies and to participants for the production of particular content or the role they might play in promoting it.

    Decentralized policymaking and moderation could create a more resilient ecosystem for free expression while allowing users to address harmful or violative content at the individual or community level. When users and communities have the power to set norms, establish policies, and make moderation decisions—rather than relying on centralized corporate or government control that applies universally—the stakes are lowered. This, in turn, may reduce tension and polarization, diminishing the incentive to engage in what \textcite{diresta} described as “working the refs.” The tendency to “work the refs” in order to influence universal or platform-wide rules and enforcement favoring (or prohibiting) specific groups, viewpoints or agendas have fueled an ever-growing supply of private and governmental platform regulations that restrict the practical exercise of free speech. Decentralized, community-led systems could mitigate this overreach and the gradual expansion of restrictive policies by allowing users to establish moderation rules that reflect their community’s norms \cite{masnick}. Such models might thus help preserve free expression without undue influence from top-down regulatory mandates, which often cause significant collateral damage to lawful speech and fuel polarizing, intractable disputes over where to draw the line—as we’ve seen throughout this discussion.

\subsection{Governance}

    Invoking the analogy between federal government and accountability naturally suggests not just a pathway for sharing accountability downwards through subsidiarity, but also principles for how communities might together influence the overall governance of the platform.  Platforms generally seek extensive feedback from participants on platform design and policies.  In some cases these proceed through ``user research'' and surveys, but in more cooperative or public interest contexts it may extend to real power over platform decisions, including formal representation \cite{ovadya2024}.  In either context, the appropriate way to synthesize the views of diverse stakeholders is a critical challenge.

    Key principles in many federal and confederal governance structures are \textit{local representation}, where stakeholders are represented by members that are part of their subsidiary jurisdiction, and \textit{degressive proportionality}, where larger subgroups have greater but not proportionally greater representation \cite{weyl2022pluralist}.  Degressive proportionality play a key role in the design of our concept of bridging scores as well as how they are aggregated into feed scoring: bridging, for example, directly incorporates the idea of degressive proportionality by less-than-proportionately counting increased support within a subcommunity compared to across communities.  Similarly, we aim to explicitly incorporate local representation through direct relationships with community leaders. In this sense, the quantitative, algorithmic governance of our system already incorporates principles of federal governance.

    However, it obviously applies these in a high throughput and highly quantitative way.  For input and governance decisions on broader platform design and policy-setting, as explored by \textcite{ovadya2023b}, it might be natural to apply similar principles in the context of deliberation, seeking overrepresentation of smaller ``minority'' groups in discussions, incorporating formal representative structures of communities and explicitly accounting for leading preexisting population divides as in many sampling procedures \cite{flanigan2021}, but using categories identified by on-platform data as in our community definition protocols.  While these could simply be used for market research purposes, they also open up the possibility of harnessing a social media system as a substrate for a more ambitious and creative form of \pluralism governance.

\subsection{Exploration}

    We suggest above using a system of community-specific ``reputation'' or ``stake'' for citizens to seed new content prior to feedback from fellow citizens.  While we believe such a reputation system is critical both for exposing communities to content likely to benefit them and encouraging good behavior by content creators, it obviously creates within-community inequality and a barrier to entry for new community members.  One way we briefly discuss above to overcome this is to allow citizens to provide verification of aspects of the identity and relationship to the community that establish unique humanity and external community standing to receive an initial digital standing within the on-platform community.  However, even this is likely insufficient to ensure citizens are exposed to an optimally diverse array of content, allowing them to explore potential unexpectedly valuable content and allowing previously marginal content creators to rise in standing.

    Natural approaches to overcoming this challenge could apply a mixture of ideas from the large literature in computer science on the ``bandit problem'' of exploration v. exploitation \cite{lattimore2020} and literature in economics on redistributive taxation \cite{diamond2011,guvenen2023}, applied to within-community reputation.  Such an approach could conserve the agency of citizens to choose which content to promote in what communities and thus what kind of reputation to stake while at the same time ensuring sufficient effective circulation of such reputation to new community entrants or those whose reputations have declined to allow a vibrant mix of exploration encouragement of good behavior and content.

\subsection{Social Infrastructure for an Automated World}

    The emergence of increasingly capable artificial intelligence systems, particularly large language models and autonomous agents, portends fundamental changes to economic and social organization that may be comparable in scale to the industrial revolution \cite{frey2017}. Unlike previous technological transitions focused primarily on physical labor, automation of cognitive work through AI threatens to disrupt the very mechanisms through which societies achieve consensus and make collective decisions \cite{couldry2024}. Our proposed design offers potential infrastructure for maintaining human agency and flourishing in such a transformed environment.
    
    The technical architecture we describe could be extended to create what \cite{bommasani2022} term ``hybrid intelligence systems'' --- not by treating AI systems as citizens, but rather by using them as tools to amplify human communities' capacity for bridging and consensus-building. The critical innovation would be using the hypergraph structure and bridging metrics to ensure AI capabilities remain oriented toward strengthening rather than supplanting human social fabric. Foundation models could be specialized for detecting and surfacing latent consensus across community divisions \cite{huang2024}, while remaining explicitly subordinate to human communities' standards and values.
    
    This addresses a fundamental challenge identified by \cite{acemoglu2022}: as automation increases, maintaining human agency requires not just economic reorganization but new mechanisms for collective choice and social coordination. Traditional democratic institutions evolved in an environment of relatively symmetric human capabilities. In a world of pervasive automation and AI assistance, institutions must be explicitly designed to preserve human social capital rather than allowing it to atrophy through disuse \cite{arrieta2018}.
    
    The economic model becomes particularly vital in this context. Rather than simply monetizing attention, the system we propose creates explicit markets for social cohesion and consensus-building --- critical resources that \cite{lanier2018} argue will become increasingly scarce as automation progresses. Communities could direct automated systems toward strengthening their internal fabric, creating what \cite{tang2016} terms ``prosocial automation'' focused on enhancing rather than replacing human meaning-making capabilities.
    
    Recent experiments like ``broad listening'' systems \cite{konya2023} and collective constitutional AI \cite{huang2024} hint at this potential, but remain limited by their reliance on existing social media architectures optimized for engagement rather than cohesion. Our proposal offers a more fundamental redesign oriented toward ``democratic infrastructure'' --- systems that systematically amplify human collective intelligence rather than attempting to automate it away.
    
    This application illuminates the deeper significance of \pluralism in an automated future: not merely as a philosophical framework but as practical infrastructure for maintaining human flourishing as automation transforms the economic basis of society. It suggests a path toward what \cite{hamdy2024} term ``plurality-preserving progress'' --- technological advancement that strengthens rather than erodes human social fabric and agency.

\section{Limitations and Risks}
\label{sec:limitations}

    While the design we propose aims to address many of the challenges of current social media environments, it is a speculative and untested system that comes with significant limitations and potential risks that require further research and careful consideration. We outline some of the key areas of concern below.

\paragraph{General speculativeness and lack of real-world testing}

    Perhaps the most obvious and overarching limitation of our proposed design is that it is largely speculative at this stage and has not been implemented and tested in real-world social media environments. There are open questions around the viability of several components, including:
    \begin{itemize}
        
        \item whether \textit{technologies} for clustering and community detection can be deployed at scale with sufficient accuracy and robustness for the use cases we describe, with appropriate privacy guarantees and, where as applicable, be integrated with existing platform backends and open protocols.
        
        \item whether the \textit{business model} of charging communities for coherence and citizens for integrity will generate sufficient value (and sufficiently \textit{legible} value) to be commercially viable. While we have argued for the benefits communities and citizens are likely to derive, it is unclear whether these will be sufficient in practice to drive adoption and ongoing payment, especially during a transition period from currently dominant models.
        
        \item whether the implied shifts in online \textit{norms, expectations, and behaviors} are attainable (e.g.,  through education and UX design), particularly given that some elements of the design ask for more active participation and initiative on the part of citizens than is the case for current platforms. Most of the more effortful affordances (e.g., ``user controls'') on current platforms are used by only small minority of users \cite{cunningham2024}.
        
    \end{itemize}
    There are additional challenges associated with transition to such a paradigm from the status quo, and while we sketch some plausible strategies in Section \ref{sec:applications}, overcoming entrenched interests and network efforts where necessary, even incrementally, is likely to be difficult.

\paragraph{Potential for unintended consequences}

    Our design aims to build social capital, empower communities and citizens, and more constructively manage social divisions. However, as with any sociotechnical system intervening in complex social processes, unintended and even contrary effects are possible. While social capital is usually described as a good, it is perhaps more accurately seen as a resource that can be harnessed support social coordination in general. For example, \textcite{satyanath2017} provide evidence that regions of Weimar Germany with higher social capital tended to be those in which the Nazi party most rapidly consolidated power. In another sense, desires to improve social media governance ties into a long line of historical efforts to envision "better societies" — some more or less positive, and others disastrous. It is the duty of any project for social improvement, sociotechnical or not, to open its ideas for public deliberation and iterate policies in concrete and evolving realities. As such, ongoing evaluation of the impacts of such a system on social dynamics, its impacts on different communities, and integration with effective governance infrastructure, is essential. One solution would be to invest in a public forum for a continuous deliberation of platform design, from general functionalities, to single features, moderation policies (where needed), or other.  

\paragraph{Challenges in evaluation and oversight}

    That acknowledged, as a pervasive sociotechnical system, our design inherits many of the challenges in rigorously evaluating existing the impacts of existing social media design choices, including difficulties of isolating experimental control and treatment groups of citizens or communities in thoroughly interconnected societies, and barriers to data access or the ability to experimentally intervene on the part of the part of impartial third-party evaluators \cite{thorburn2022,naranayan2023}. This complicates efforts towards effective oversight.

\paragraph{Legal considerations} %

    We believe our design is compatible with current regulatory environments in, e.g., the US and the EU, though flag two particularly relevant legal considerations.
    
    First, in some jurisdictions there are requirements for transparency about which organic content (created by non-paying citizens) and inorganic content (namely, ads). In our design this distinction is in some ways blurrier, as relatively more of the content that citizens see will appear in their feeds because it was valued and paid for by the communities to which they belong, or because they themselves are paying the platform for content which bridges across those same communities. Transparency about the relevant payments, to the extent they are the reason content is shown to a citizen, is important both ethically and to be compliant with such laws.

    Second, many jurisdictions have regulations  to ensure responsible use of data about individuals, with particular focus on the collection and use of ``protected attributes'' such as age, gender, ethnicity, and religion. Communities defined by such attributes are socially consequential and believe it is important that they are formally represented on the platforms we describe. Our understanding is that there are no hard legal barriers to the collection, inference or use of such attributes so long as there is appropriate transparency, informed consent, and privacy protections in place. However, more thorough legal assessments for specific implementations of this design will need to be carried out, and it's possible that some incompatibilities would emerge that would require regulatory changes before they could be deployed.

\paragraph{Inherent tensions and tradeoffs}

    Finally, it is important to recognize that many of the goals we aim to achieve exist in tension. Notably this includes bridging and balancing --- as more attention is allocated towards bridging content, the less is available for attending to diverse, conflicting perspectives, and vice versa. Similarly for complex pluralism and personal integrity --- complexity in community structure that strengthens social fabric can be disorienting and alienating for citizens, and conversely, simpler community structures that are more legible to citizens tend to weaken the social fabric. We believe that it is reasonable to aim for an equilibrium between such objectives, where none is optimized so aggressively that it substantively undermines the others.

\section{Conclusion}
\label{sec:conclusion}

    A tremendous amount has been written about the social media phenomenon, which itself grew out of a deep sociological literature.  A range of aspects of it have been optimized to fulfill the economic and social structure it has developed.  And it has been endlessly critiqued for a variety of social harms it causes.  Yet with the important partial exception of danah boyd's early work, the overall design objectives and philosophy it is intended to serve have been deeply undertheorized.  An extremely rich sociotechnical system has, in practice, been designed to serve extremely simplistic goals of individual level engagement and personalized advertising revenue maximization, with ancillary harms being addressed in one-off ways with orthogonal hammers.  In this paper we aimed to bridge the gap between the simplistic design philosophy and the richness of the system by providing a conceptual frame and connection to technical tools for a design of social media.  In our approach, social media is designed to embody \pluralism, creating social self-consciousness of the hypergraph that constitutes the social fabric and charging the communities and citizens that constitute this graph for the coherence that their slice of this self-consciousness supports.

    While we believe this ambition is important to meaningfully and comprehensively reimagining the system, we took several shortcuts to achieve that goal.  First, to capture the breadth of this goal with any kind of brevity, we have skirted over a range of more detailed questions that we hope will be the focus of future research.  These include:

    \begin{enumerate}
    
    \item Most importantly, we fall far short of providing a full specification here for a prototype system that could be directly used in experiments.  Such a prototype would require filling in a variety of setting-specific details we deliberately abstracted from.  We look forward to seeing a range of such prototypes develop and share learnings across contexts.

    \item While we abstracted from detail, we did so insufficiently to provide a full formalism that could be used to derive analytic proofs or simulations of the system to establish some of the properties we argued for in Section \ref{sec:discussion} deductively or computationally.  While it is likely infeasible to construct a single model capturing all the elements analyzed here, we hope that various simplified slices of our argument can be illuminated, amplified, qualified and improved by such analysis and look forward to that work.

    \item Especially underspecified were the design of the elaborations of the system in Section \ref{sec:elaborations}, each of which on its own likely deserves a treatment of similar length and depth to that accorded the broad system design here.  In many application contexts several of these will be critical to implementation.

    \item Another crucial aspect we neglect is measurement of the desired outcomes both generally described in Section \ref{sec:discussion} and for the specific contexts of application mapped in Section \ref{sec:applications}.  Such measurements will be important components of persuading stakeholders from communities to regulators that the system's features are worth paying for and supporting as alternatives to the structure of current social media. 

    \item Finally,  we gave almost no attention to the role of law, regulation, corporate legal structure or extra-platform social movements in bringing about and supporting the design described here.  To a significant extent this was deliberate, given that there has been far more attention devoted to these topics on the theme of social media reform than to detailed designs for an alternative model, including in much of our own previous work.  However, bridging the gap thus created between this design and the supporting social and institutional frameworks is an important goal in future work.
    
    \end{enumerate}

    Second, a more fundamental compromise is core to our design, namely that between the richness of social structure and the limitations of digital models on the other. On the one hand, while the model we develop and aim to make transparent to citizens is more sociologically grounded than most existing interfaces, it is still wildly simplistic compared either to the true richness of social life and even the most sophisticated widely appreciated sociological accounts \cite{couldry2018}.  On the other hand, there are real questions as to whether the format we describe already asks too much of many citizens, especially those who are less engaged with the platform, in making platform features work (to mention, for example, consenting to micropayments).  Our design thus does not create complete transparency, but rather make a step towards bridging the currently gaping divide between the understanding citizens have of social structure and the way understanding of this structure is used to serve them content.

    Moving beyond this first step requires significant additional work from both the vantage of political philosophy, psychology, education and assistive technology on the one hand and improved computational sociology on the other; both are crucial areas for future research. Ultimately, our work participates in ongoing efforts to translate public values into platform design. A fundamental aspect of this activity is that it is invariably open to public deliberation, as other stakeholders may have different approaches to translating one or another value into a technical function. That is the nature of techno-political deliberation, and ultimately a form of policy making in its own right. 

    In the first area, research and development of practices on sociological and sociopolitical psychology, communication, and education, as well as on assistive technology to aid in understanding social systems, are critical to try to enhance the capacity of citizens to understand social complexity and of tools to aid them in such understanding.  For example, well-documented cross-cultural comparisons, such as the work of \textcite{shields1989} and \textcite{roesgaard2016} on Japanese education, show that it is possible to create much greater social awareness in education than is typical in mainstream Western education.  \textcite{freinacht2019} emphasizes several practices common in Scandinavian societies that facilitate greater social awareness and participative design.  None of this is particularly novel: the rise of mass education, technical skills and assistive technology were critical to both the initial rise of industrialism and especially the East Asian productivity miracle inspired by Deming's vision of workers with awareness of the full production process, thereby able to continuously improve quality \cite{goldin2009race}. The survival of the planet's ecology has deepened on an environmental movement and accompanying advances in science and environmental awareness that has allowed us to detect and react to a growing range of anthropogenic threats to the planet \cite{bratton}.  Significant additional progress towards social self-consciousness will require analogous advances in sociological analogs of such practices coordinated with computational sociological systems capable of modeling the dynamics of social complexity.

    In the second area, state of the art models of human sociality (along with many other phenomena) are increasingly based on large ``black box'' models, usually with a deep neural network architecture, with most leading social media platforms employing these heavily in their recommender systems.  These models can capture and suggest much richer descriptions of social life than the comparatively simple hypergraphical model we study, allowing direct modeling of semiotic systems \cite{wedeen2002conceptualizing}, the non-linear interactions of intersectional identity \cite{crenshaw}, social power dynamics \cite{bourdieu1984espace} and much more. Yet the common interpretation of these models is as autonomous intelligent agents, rather than as models or components of the social fabric, contrary to pioneering figures in the field who viewed them in politico-economic terms \cite{holland,carley1995computational}.  More cybernetic interpretations and applications of such networks as representing and embedding human relationships, including according economic and political rights to those contributing to the networks, will be crucial to bridging the socio-technical gap from the technical side.  Only by making people more aware of their social situation and understanding large models as maps of such situations can we hope to fully harness the capability of technology to create the collective social self-awareness that \textcite{lovelock1983gaia} articulated as central to the ``Gaia paradigm'', as well as a technological future consistent with many faith-based transcendent value systems \cite{hamdy}.

\clearpage

\section*{Funders}

    Luke Thorburn was supported in part by UK Research and Innovation [grant number EP/S023356/1], in the UKRI Centre for Doctoral Training in Safe and Trusted Artificial Intelligence (safeandtrustedai.org), King's College London.

\nocite{*}
\printbibliography

\end{document}